\documentclass[9pt,twocolumn,twoside]{article}
\usepackage{fancyhdr}
\usepackage{amssymb}
\usepackage[ruled,vlined,algo2e]{algorithm2e}
\usepackage{algorithm}
\include{pythonlisting}
\usepackage{times}
\usepackage{graphicx}
\usepackage{multirow}
\usepackage{diagbox}
\usepackage{adjustbox}
\usepackage{amsthm}
\usepackage{float}
\usepackage{dsfont}
\usepackage{amsmath}
\usepackage{booktabs}
\usepackage{amsfonts}
\usepackage{enumitem}
\usepackage{xcolor}
\usepackage{hyperref}

\newcommand{\minimize}{\mathop{\mathrm{minimize}}}

\newcommand{\subjectto}{\mbox{subject to}}

\newcommand{\ones}{\mathds{1}_n}
\newcommand{\bigo}{\mathcal{O}}
\newcommand{\bigoprob}{\mathcal{O}_\mathbb{P}}
\newcommand{\sumtot}{\sum\limits_{t=1}^T}
\newcommand{\indicator}{\mathds{1}}
\newcommand{\sumtwotot}{\sum\limits_{t=2}^T}
\newcommand{\halft}{\frac{1}{2T}}
\newcommand{\leqwithspaces}{\;\;\leq\;\;}

\usepackage{xcolor}
\usepackage{xparse,xcoffins}

\ExplSyntaxOn
\NewCoffin\imagecoffin
\NewCoffin\labelcoffin

\keys_define:nn { omid/label }
{
    label   .tl_set:N = \l_omid_label_tl,
    labelbox .bool_set:N = \l_omid_label_box_bool,
    labelbox .default:n = true,
    fontsize .tl_set:N = \l_omid_label_size_tl,
    fontsize .initial:n = \footnotesize,
    pos .choice:,
    pos/nw .code:n = \tl_set:Nn \l_omid_label_pos_tl { left,up },
    pos/ne .code:n = \tl_set:Nn \l_omid_label_pos_tl { right,up },
    pos/sw .code:n = \tl_set:Nn \l_omid_label_pos_tl { left,down },
    pos/se .code:n = \tl_set:Nn \l_omid_label_pos_tl { right,down },
    pos/n .code:n = \tl_set:Nn \l_omid_label_pos_tl { hc,up },
    pos/w .code:n = \tl_set:Nn \l_omid_label_pos_tl { left,vc },
    pos/s .code:n = \tl_set:Nn \l_omid_label_pos_tl { hc,down },
    pos/e .code:n = \tl_set:Nn \l_omid_label_pos_tl { right,vc },
    pos .initial:n = nw,
    unknown .code:n   = \clist_put_right:Nx \l_omid_label_clist
                       { \l_keys_key_tl = \exp_not:n { #1 } }
}
\clist_new:N \l_omid_label_clist
\box_new:N \l_omid_label_box
\box_new:N \l_omid_label_image_box

\NewDocumentCommand{\xincludegraphics}{O{}m}
{
  \group_begin:
  \tl_clear:N \l_omid_label_tl
  \clist_clear:N \l_omid_label_clist
  \keys_set:nn { omid/label } { #1 }
  \tl_if_empty:NTF \l_omid_label_tl
   {
    \omid_includegraphics:Vn \l_omid_label_clist { #2 }
   }
   {
    \SetHorizontalCoffin\imagecoffin
     {
      \omid_includegraphics:Vn \l_omid_label_clist { #2 }
     }
    \SetHorizontalCoffin\labelcoffin
     {
      \raisebox{\depth}
       {
        \bool_if:NTF \l_omid_label_box_bool
         { \fcolorbox{white}{white}{\l_omid_label_size_tl\l_omid_label_tl} }
         { \l_omid_label_size_tl\l_omid_label_tl }
       }
     }
    \SetVerticalPole\imagecoffin{left}{3pt+\CoffinWidth\labelcoffin/2}
    \SetVerticalPole\imagecoffin{right}{\Width-3pt-\CoffinWidth\labelcoffin/2}
    \SetHorizontalPole\imagecoffin{up}{\Height-3pt-\CoffinHeight\labelcoffin/2}
    \SetHorizontalPole\imagecoffin{down}{3pt+\CoffinHeight\labelcoffin/2}
    \use:x{\JoinCoffins\imagecoffin[\l_omid_label_pos_tl]\labelcoffin[vc,hc]} 
    \TypesetCoffin\imagecoffin
   }
   \group_end:
}
\NewDocumentCommand{\setlabel}{m}
{
  \keys_set:nn { omid/label } { #1 }
}

\cs_new_protected:Nn \omid_includegraphics:nn
{
  \includegraphics[#1]{#2}
}
\cs_generate_variant:Nn \omid_includegraphics:nn { V }

\ExplSyntaxOff

\RequirePackage[utf8]{inputenc}
\RequirePackage[english]{babel}
\RequirePackage{amsmath,amsfonts,amssymb}
\RequirePackage{lmodern}
\RequirePackage[scaled]{helvet}
\RequirePackage[T1]{fontenc}
\RequirePackage{lettrine} 

\RequirePackage{afterpage}
\RequirePackage{ifpdf,ifxetex}
\ifpdf\else
  \ifxetex\else
    
    \pdfpagewidth=\paperwidth
    \pdfpageheight=\paperheight
\fi\fi
\RequirePackage{xcolor}
\RequirePackage{tikz}
\RequirePackage[framemethod=tikz]{mdframed}

\newcommand{\footerfont}{\normalfont\sffamily\fontsize{7}{9} \selectfont}
\newcommand{\titlefont}{\fontfamily{lmss}\bfseries\fontsize{22pt}{24pt}\selectfont}
\newcommand{\dropcapfont}{\fontfamily{lmss}\bfseries\fontsize{26pt}{28pt}\selectfont}
\newcommand{\datesfont}{\normalfont\sffamily\fontsize{7}{8}\selectfont}
\newcommand{\absfont}{\normalfont\sffamily\bfseries\fontsize{8}{11}\selectfont}
\newcommand{\keywordsfont}{\normalfont\rmfamily\fontsize{7}{10}\selectfont}

\RequirePackage{authblk}
\renewcommand\Affilfont{\color{color0}\normalfont\sffamily\fontsize{7}{8}\selectfont}
\makeatletter
\renewcommand\AB@affilsepx{; \protect\Affilfont}
\makeatother

\RequirePackage{xifthen}
\newboolean{shortarticle}
\newboolean{singlecolumn}

\RequirePackage{graphicx,xcolor}
\RequirePackage{colortbl}
\RequirePackage{booktabs}
\RequirePackage[noend]{algpseudocode}
\RequirePackage{changepage}
\RequirePackage[twoside,%
        letterpaper,includeheadfoot,%
        layoutsize={8.125in,10.875in},%
                layouthoffset=0.1875in,%
                layoutvoffset=0.0625in,%
                left=38.5pt,%
                right=43pt,%
                top=43pt,
                bottom=32pt,%
                headheight=0pt,
                headsep=10pt,%
                footskip=25pt]{geometry}
\RequirePackage[labelfont={bf,sf},%
                labelsep=period,%
                figurename=Fig.]{caption}
\definecolor{black50}{gray}{0.5} 
\definecolor{color0}{RGB}{0,0,0} 
\definecolor{color1}{RGB}{59,90,198} 
\definecolor{color2}{RGB}{16,131,16} %
\definecolor{templbluetext}{RGB}{0,101,165} %
\definecolor{templblueback}{RGB}{205,217,235} %
\RequirePackage[numbers,sort&compress,merge,round]{natbib}
{
    {
    \bibliographystyle{unsrtnat}
  }
}

\makeatletter 
\renewcommand\@biblabel[1]{ #1.} 
\def\tagform@#1{\maketag@@@{\bfseries(\ignorespaces#1\unskip\@@italiccorr)}}
\renewcommand{\eqref}[1]{\textup{{\normalfont Eq.~(\ref{#1}}\normalfont)}}
\makeatother

\DeclareCaptionFormat{templformat}{\normalfont\sffamily\fontsize{7}{9}\selectfont#1#2#3}
\RequirePackage{etoolbox}
\AtBeginEnvironment{tabular}{
\sffamily\fontsize{7.5}{10}\selectfont
}

\makeatletter
\renewcommand\tagform@[1]{\maketag@@@ {[\ignorespaces #1\unskip \@@italiccorr ]}}
\makeatother

\RequirePackage{fancyhdr}  
\RequirePackage{lastpage}  
\makeatletter
\fancypagestyle{firststyle}{
   \fancyfoot[R]{\footerfont \hspace{7pt}|\hspace{7pt}\textbf{\today}\hspace{7pt}|\hspace{7pt}vol. XXX\hspace{7pt}|\hspace{7pt}no. XX\hspace{7pt}|\hspace{7pt}\textbf{\thepage\textendash\pageref{LastPage}}}
   \fancyfoot[L]{\footerfont\@ifundefined{@doi}{}{\@doi}}
}
\makeatother

\makeatletter
\RequirePackage[explicit]{titlesec}
\renewcommand{\thesubsection}{\Alph{subsection}}

\titleformat{\section}
  {\large\sffamily\bfseries}
  {\thesection.}
  {0.5em}
  {#1}
  []
\titleformat{name=\section,numberless}
  {\large\sffamily\bfseries}
  {}
  {0em}
  {#1}
  []  
\titleformat{\subsection}[runin]
  {\sffamily\bfseries}
  {\thesubsection.}
  {0.5em}
  {#1. }
  []
\titleformat{\subsubsection}[runin]
  {\sffamily\small\bfseries\itshape}
  {\thesubsubsection.}
  {0.5em}
  {#1. }
  []    
\titleformat{\paragraph}[runin]
  {\sffamily\small\bfseries}
  {}
  {0em}
  {#1} 
\titlespacing*{\section}{0pc}{3ex \@plus4pt \@minus3pt}{5pt}
\titlespacing*{\subsection}{0pc}{2.5ex \@plus3pt \@minus2pt}{2pt}
\titlespacing*{\subsubsection}{0pc}{2ex \@plus2.5pt \@minus1.5pt}{2pt}
\titlespacing*{\paragraph}{0pc}{1.5ex \@plus2pt \@minus1pt}{12pt}

\newcommand{\additionalelement}[1]{\def\@additionalelement{#1}}
\newcommand{\dates}[1]{\def\@dates{#1}}
\newcommand{\doi}[1]{\def\@doi{#1}}
\newcommand{\leadauthor}[1]{\def\@leadauthor{#1}}
\newcommand{\etal}[1]{\def\@etal{#1}}
\newcommand{\keywords}[1]{\def\@keywords{#1}}
\newcommand{\authorcontributions}[1]{\def\@authorcontributions{#1}}
\newcommand{\authordeclaration}[1]{\def\@authordeclaration{#1}}
\newcommand{\equalauthors}[1]{\def\@equalauthors{#1}}
\newcommand{\correspondingauthor}[1]{\def\@correspondingauthor{#1}}
\newcommand{\significancestatement}[1]{\def\@significancestatement{#1}}
\newcommand{\matmethods}[1]{\def\@matmethods{#1}}
\newcommand{\acknow}[1]{\def\@acknow{#1}}

\newcommand{\dropcap}[1]{\lettrine[lines=2,lraise=0.05,findent=0.1em, nindent=0em]{{\dropcapfont{#1}}}{}}

\def\xabstract{abstract}
\long\def\abstract#1\end#2{\def\two{#2}\ifx\two\xabstract 
\long\gdef\theabstract{\ignorespaces#1}
\def\go{\end{abstract}}\else
\typeout{^^J^^J PLEASE DO NOT USE ANY \string\begin\space \string\end^^J
COMMANDS WITHIN ABSTRACT^^J^^J}#1\end{#2}
\gdef\theabstract{\vskip12pt BADLY FORMED ABSTRACT: PLEASE DO
NOT USE {\tt\string\begin...\string\end} COMMANDS WITHIN
THE ABSTRACT\vskip12pt}\let\go\relax\fi
\go}

\newcommand{\abscontent}{
\noindent
{%
\parbox{\dimexpr\linewidth}{%
    \vskip3pt%
  \absfont \theabstract
}%
}%
\vskip10pt%
\noindent
{\parbox{\dimexpr\linewidth}{%
{
 \keywordsfont \@ifundefined{@keywords}{}{\@keywords}}%
}}%
\vskip12pt%
}

\newlength\templ@vertadjust
\newcommand\verticaladjustment[1]{\setlength{\templ@vertadjust}{#1}}

\renewcommand{\@maketitle}{%
{%
\ifthenelse{\boolean{shortarticle}}
{\ifthenelse{\boolean{singlecolumn}}{}{
{\raggedright\baselineskip= 24pt\titlefont \@title\par}%
\vskip10pt
{\raggedright \@author\par}
\vskip8pt
{\raggedright \datesfont \@ifundefined{@dates}{}{\@dates}\par}
\vskip12pt%
}}
{
\vskip10pt%
{\raggedright\baselineskip= 24pt\titlefont \@title\par}%
\vskip10pt
{\raggedright \@author\par}
\vskip8pt
{\raggedright \datesfont \@ifundefined{@dates}{}{\@dates}\par}
\vskip12pt
{%
\abscontent
}%
\vskip25pt%
}%
}%
\vskip\templ@vertadjust
}%

\RequirePackage[flushmargin,ragged,symbol*]{footmisc}
\renewcommand{\footnoterule}{
  \kern -3pt
  {\color{black50} \hrule width 72pt height 0.25pt}
  \kern 2.5pt
}

\titleclass{\acknow@section}{straight}[\part]
\newcounter{acknow@section}
\providecommand*{\toclevel@acknow@section}{0}
\titleformat{\acknow@section}[runin]
   {\sffamily\normalsize\bfseries}
   {}
   {0em}
   {#1.}
   []
\titlespacing{\acknow@section}
  {0pt}
  {3.25ex plus 1ex minus .2ex}
  {1.5ex plus .2ex}

\newcommand{\showacknow}{
\@ifundefined{@acknow}{}{
\vskip 3.25ex plus 1ex minus .2ex
\noindent{\sffamily\normalsize\bfseries ACKNOWLEDGMENTS.\hspace{1.5ex plus .2ex}}
\small\@acknow}
}

\RequirePackage{enumitem} 

\RequirePackage[rightcaption]{sidecap}

\graphicspath{{./Figures/}}

\title{The 1995-2018 Global Evolution of the Network of Amicable and Hostile Relations Among Nation-States}

\author[a]{Omid Askarisichani}
\author[a]{Ambuj K. Singh}
\author[b,c]{Francesco Bullo}
\author[b,d]{Noah E. Friedkin}

\affil[a]{Department of Computer Science, University of California, Santa Barbara, CA 93106}
\affil[b]{Center for Control, Dynamical Systems and Computation, University of California, Santa Barbara, CA 93106}
\affil[c]{Department of Mechanical Engineering, University of California, Santa Barbara, CA 93106}
\affil[d]{Department of Sociology, University of California, Santa Barbara, CA 93106}

\leadauthor{askarisichani}

\authordeclaration{The authors declare no conflicts of interest.}
\correspondingauthor{\textsuperscript{*}To whom correspondence should be addressed. E-mail: oaskarisichani\@cs.ucbs.edu}

\keywords{international news $|$ balance theory $|$ quantitative sociology $|$ convex optimization $|$ time-varying markov $|$ data mining $|$ international conflicts}

\begin{abstract}
We draw on the data collected by the Integrated Crisis Early Warning System on millions of international and regional public news stories, and this system's indicators of the orientation toward a specific nation-state. We construct the networks of international amicable and hostile relations among nation-states that occur in specific time-periods in order to study the global evolution of the network of such international appraisals. Our analysis presents evidence of an evolution of the structure of this network and a model of the probabilistic micro-dynamics of the alterations of international appraisals during the 1995-2018 span of the available data. Our research provides empirical findings on long-standing debates in the interdisciplinary field of work on Structural Balance Theory. Also remarkably, we find that the trajectory of the Frobenius norm of sequential transition probabilities, which govern the evolution of international appraisals among nations, dramatically stabilizes.
\end{abstract}

\dates{This manuscript was compiled on \today}

\begin{document}

\verticaladjustment{-2pt}

\maketitle
\thispagestyle{firststyle}
\ifthenelse{\boolean{shortarticle}}{\ifthenelse{\boolean{singlecolumn}}{\abscontentformatted}{\abscontent}}{}

\dropcap{T}he Integrated Crisis Early Warning System (ICEWS) is a comprehensive, automated, and validated system to monitor national, sub-national, and internal crises. Its event data is publicly available and consists of coded interactions between socio-political actors (i.e., friendly or hostile actions between individuals, groups, sectors, and nation-states). Geographical-temporal metadata are extracted and associated with the relevant events within a news article. The data structure is a list of events. Every event has an occurrence date, a source actor, and a target actor. Every event is also annotated with a value in the $[-10, +10]$ interval that indicates the orientation of the source to the target actor: $-10$ (completely offensive) to $+10$ (completely supportive). For instance, the news event ``Japan said on Tuesday it had halted economic aid to Yugoslavia in line with Western efforts to end the fighting there'' is coded as a directed edge from Japan to Yugoslavia with weight $-5.6$, that is calculated based on the content of the news and the type of event (which in this case was "reduce or stop economic assistance"). Generally, the actors have political positions in a particular country, such as government administration, military, police, etc. In our analysis, we consider every country as a node and focus our analysis on international events in which the source and target nodes belong to different countries. The data includes 250 countries (network nodes) and over 8 million international events (network edges) occurring over more than two decades.

These data provide a unique opportunity to (i) construct networks of international amicable and hostile relations among nation-states that occur in specific time-periods and (ii) investigate the global evolution of the network of such international appraisals over a lengthy span of time. The motivations for exploiting such data include an understanding of the origins of war, the formation of alliances, and the balance of powers. Similarly motivated research includes~\cite{jackson2015networks, zheng2015social, mcdonald1985alliance, easley2012networks, harary1961structural, doreian2015structural, belaza2017statistical}. Some of this research on international appraisals has been guided by a network science theory of structural balance~\cite{heider1946attitudes, cartwright1956structural} in which signed networks evolve toward either a network of all positive appraisals or a network composed of two components of actors with all positive within-component appraisals and all negative between-component appraisals. Gellman \emph{et al.}~\cite{gellman1989elusive} used structural balance theory to analyze the origins of WWI, and Antal \emph{et al.}~\cite{antal2006social} similarly used balance theory to explore the evolution of major changes among the protagonists prior to WWI during the period 1872 to 1907. Moore \emph{et al.}~\cite{moore1978international} used balance theory to analyze the conflict over Bangladesh's separation from Pakistan in 1972. Harary \emph{et al.}~\cite{harary1961structural} also analyzed international relations among nations and different states of equilibrium and disequilibrium, using structural balance theory for the crisis in the Middle East in 1956. Harary \emph{et al.}~\cite{harary1961structural} showed how ten countries, after each international shock, sought a new equilibrium alignment consistent with what balance theory predicts.

With the ICEWS data, for the first time in a longitudinal setting, we are able to address three important limitations of the line of research on the evolution of international appraisals have has been motivated by structural balance theory.

First, unlike balance theory's prediction~\cite{cartwright1956structural}, the empirical evidence does not support the prediction that a network of friends and enemies must evolve either to a network of all friends or to a network of composed two antagonistic components of actors with all positive within-component appraisals and all negative between-component appraisals. Instead, the evidence supports the conclusion that the evolution of appraisals is mainly driven by reductions of intransitive relations among actors, which allow the emergence of complex network topologies with more than two mutually antagonistic sets of countries and hierarchically structured positive relations between countries~\cite{friedkin2019positive}. Intransitive relations occur when there is evidence of a positive chain of international relations $i \xrightarrow{+} k \xrightarrow{+} j$ and evidence of a negative $i \xrightarrow{-} j$ relation. Such intransitive relations are assumed to be sources of international tensions that lead to transformations of positive relations to negative relations, and vice versa.

Second, the empirical evidence does not support balance theory's assumption that every actor has either and positive or negative orientation to every other actor, and a line of research has developed that relaxes this assumption~\cite{cartwright1956structural, harary1959measurement, abell1968structural, de1999sign, kunegis2010spectral, terzi2011spectral, facchetti2011computing, easley2012networks, rawlings2017structural}. In large-scale networks, incomplete networks that include indifference relations are the rule. In the ICEWS data, such null relations appear when there are neither amicable nor hostile events between two countries. While it may be assumed that all countries are aware of each other's existence, such awareness need not be coupled with an amicable or hostile relation.

Third, despite numerous theoretical advancements and empirical studies on balance theory, the dynamic predictions of network state changes have rarely been tested with empirical investigations of longitudinal data~\cite{szell2010, nc2019}. There have been a large number of empirical studies on static networks~\cite{leskovec2010predicting, newcomb1956prediction, shahriari2016sign}. Longitudinal studies have been limited to small populations of actors and to a small number of temporal states of the network~\cite{zheng2015social, lazer2009computational}. In contrast, this study presents results on the most extensive set of longitudinal data yet assembled that allows research on the question of whether the evolution of appraisal networks is mainly driven by reductions of intransitive relations.

Our contributions in this study are as follows. In this article, we advance the line of research on the evolution of the network of amicable and hostile relations among countries, and also the basic science on structural balance theory. To the best of our knowledge, this article reports novel empirical findings from the largest longitudinal data yet assembled on structural balance theory. First, we address the existing lacuna on balance theory dynamics in large-scale networks that include null (indifference) relations. In networks that include large numbers of null relations, we find startling evidence of changes in international relations that are predominately restricted to only 10 types of configurations in the possible set of 138 configurations of null, positive, or negative relations among any three countries. Second, we find surprising evidence that does not comport with balance theory's prediction of a general tendency toward configurations of international relations that do not violate the theory's assumptions. Instead, we find a trajectory that involves a short period of increasing numbers of violations of balance theory's expectations, as indifference relations convert to negative or positive relations, followed by a longer trajectory that involves decreasing numbers of violations of transitive relations. During the entire course of this evolution, we find that balanced triads are likely to stay balanced and that unbalanced triads are likely to transition to balanced ones. Third, we introduce a novel convex optimization model with a convergence guarantee for quantitatively estimating time-varying Markov chains of the transitions of the structure of international relations. Empirical Markov transition matrices show diminishing variability over our longitudinal data and emergent dynamic stability. Fourth, we conclude with evidence suggesting that the evolution of the network structure toward dynamic stability is subject to disturbances that appear to be related to disruptive international events and changes in the global economy. This finding provides a novel empirical support on a longitudinal setting for earlier research regarding the effect of global trades on international conflicts~\cite{jackson2015networks}.


\section{Results}
In the span of 23+ years from 1995-01-01 to 2018-09-30, the ICEWS data includes 250 countries and 8,073,921 international events, each of which are positive or negative appraisals generated from a source country to some other target country. Each appraisal is a value in the interval $[-10, +10]$ that indicates the orientation of the source to the target actor in a news event that occurred at particular time: $-10$ (completely offensive) to $+10$ (completely supportive)~\cite{shilliday2012data}. The event data present positive, negative, and null international edges. The null edges are either a source-target news event that cannot be given a sign or an indicator that no source-target news event has been published. Precisely, there are 5,974,283 positive international edges (74\%), 1,333,646 negative international edges (17\%), and null instances of 765,992 neutral edges (with weight zero) (9\%).

\subsection{Empirical Dynamic Networks}\hfill\\
To investigate the evolution of the international appraisals, these data are disaggregated into time periods. Each time period is associated with the subset of published events that occurred during the period. Each network is comprised of 62,250 ($250^2 - 250$) directed positive, negative, and null edges among the 250 countries. A particular source-target ordered pair of countries may be associated with multiple events during a particular time period, and we take the sign of the summed value of these multiple events as the measure of the orientation of the source to the target. Hence, for 3 month periods, we have 101 snapshots of signed and directed networks among countries (see section Materials and Methods). We attend to different definitions of period length as a check on the robustness of our findings.

\begin{figure}[!ht]
    \centering
    \xincludegraphics[width=0.48\linewidth, label=\textbf{a)}]{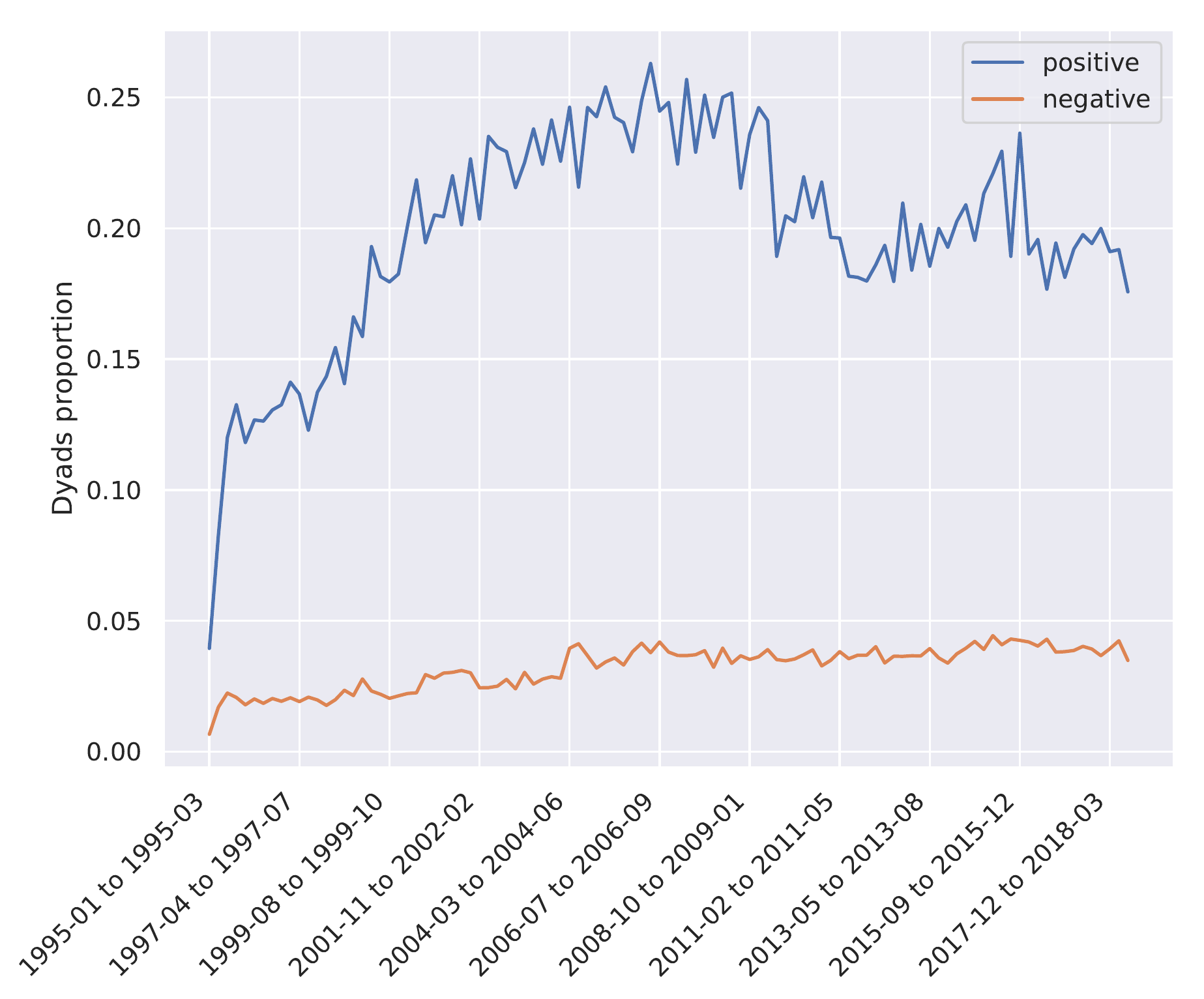}
    \xincludegraphics[width=0.48\linewidth, label=\textbf{b)}]{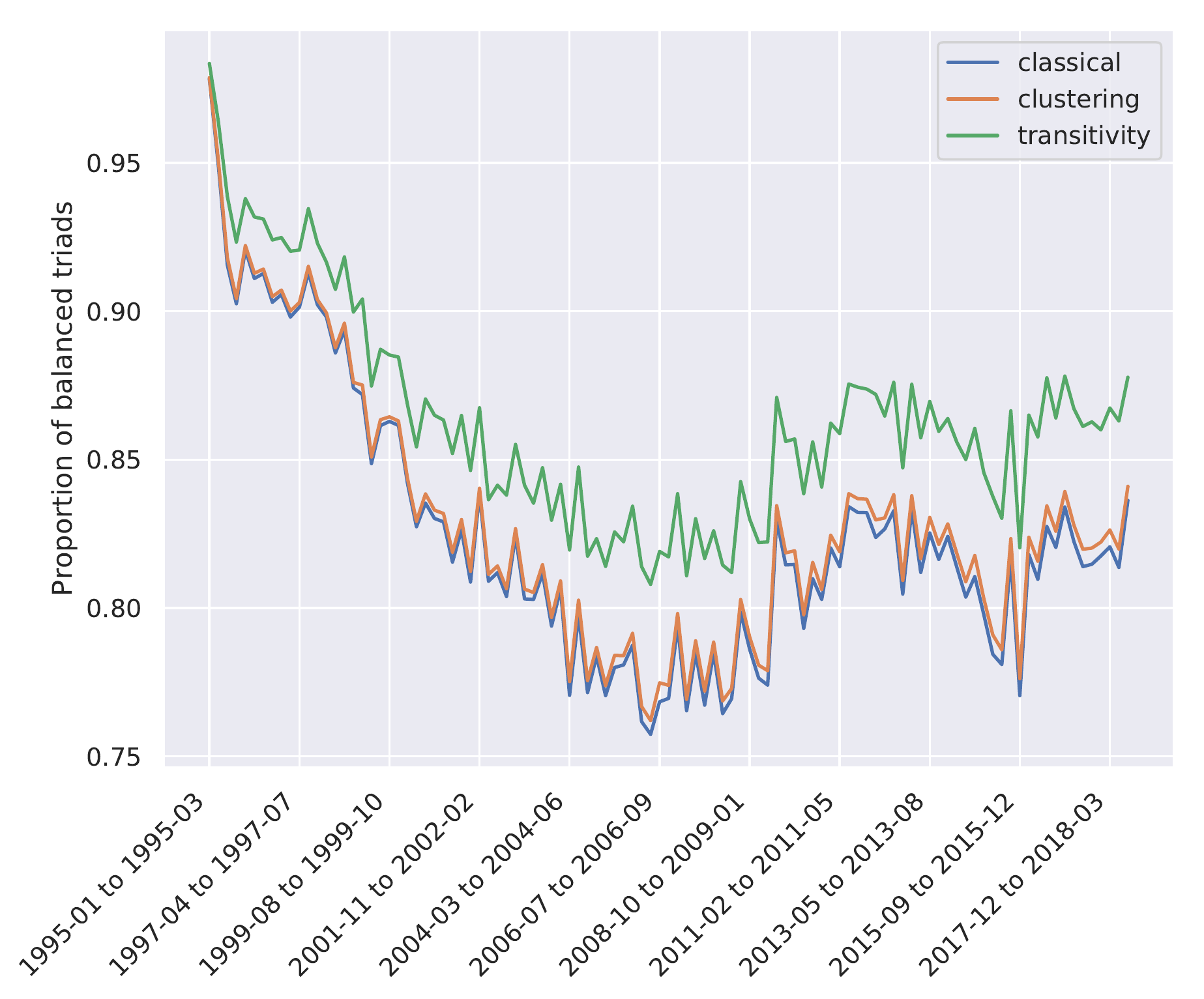}
    \xincludegraphics[width=0.48\linewidth, label=\textbf{c)}]{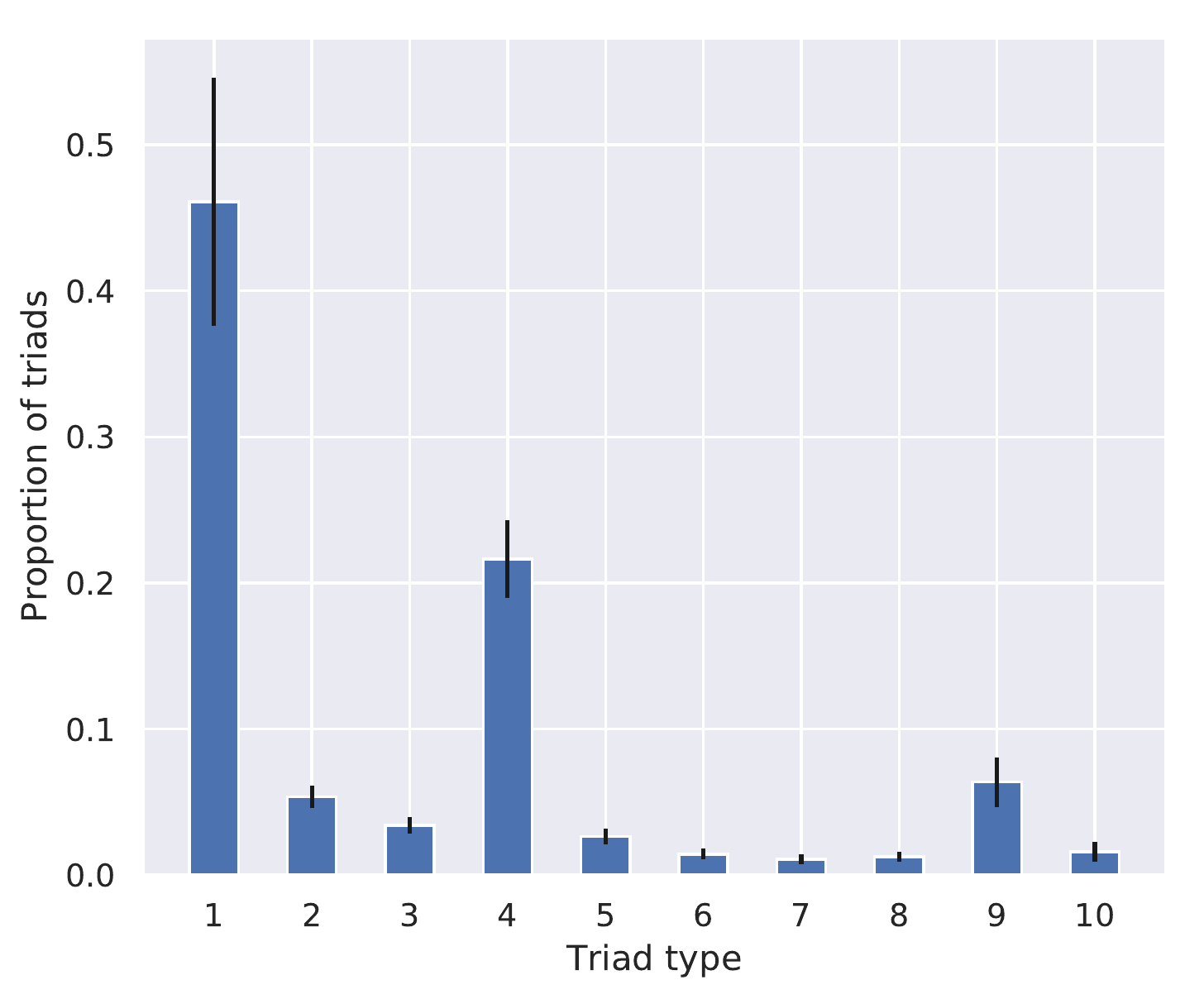}
    \xincludegraphics[width=0.8\linewidth, label=\textbf{d)}]{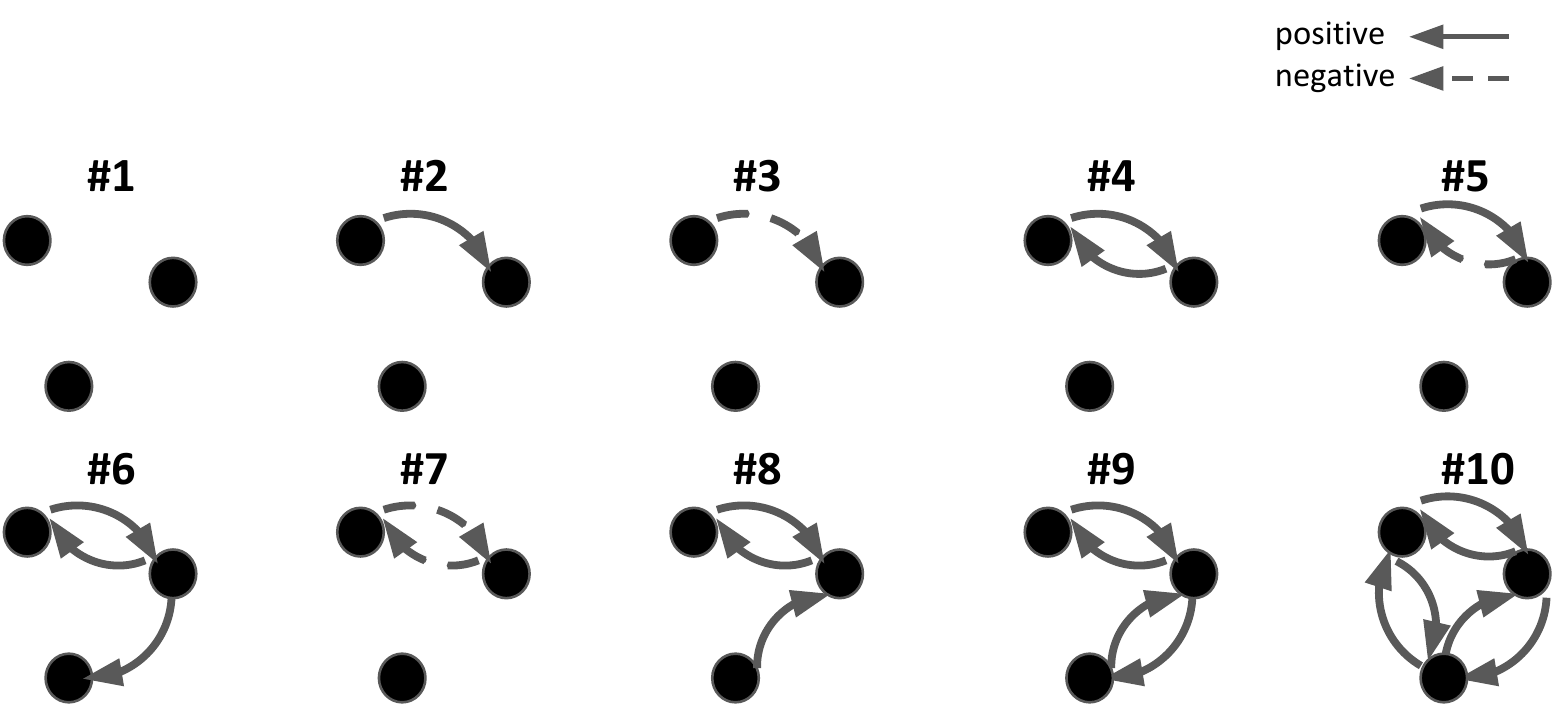}
    \caption{a) Proportion of positive and negative dyads (edges) over time in the strongly connected component. b) Proportion of balanced triads over 23+ years (with all definitions of balance in section Generalized Structural Balance. Triads are counted in networks extracted by aggregating three months of news articles. We find changing the period length to monthly, biweekly and weekly does not change the trend in this figure. c) Average and standard deviation of proportion of triads over 23+ years. d) The set of 10 operative triads (out of 138 possible triads).}
\label{fig:icews_triad_distribution}
\end{figure}

\paragraph{Core-periphery phenomenon:}
We find that the network structures are in the class of classical core-periphery (also called center-periphery) structures~\cite{shils1975center, bourgeois2001distant}. Such structures of $n$ nodes are composed of one strong component of $k$ nodes (the core) in which one or more paths of positive appraisals exits from every member of the core to every other member of the core. The remaining set of $n-k$ nodes (the periphery) is composed of nodes each which has at least one positive appraisal to a node (or nodes) in the core.  
We find, in the first period of the data, that there are exactly 111 countries in the core and 23 countries in the periphery. These peripheral countries were Afghanistan, Angola, Guinea, Haiti, Sierra Leone, Zimbabwe, Bolivia, Paraguay, Rwanda, Armenia, Azerbaijan, Congo, Grenada, Guatemala, Guyana, Kuwait, Malawi, Mozambique, Nicaragua, Nigeria, Panama, Sudan, Timor-Leste. We find that over time (in about 4 years), all of the first period peripheral countries moved into the core.  That is, the size of the core grew to $n=134$ and was maintained as a single strong component of 134 nodes. The remaining 116 countries do not exist in all network periods. Thus, we focus on the network dynamics of these 134 countries in our analysis. Using quarterly periods, the percentage of positive ties in the core increases from 4\% to 18\%, and the percentage of negative ties increases from 1\% to 4\%. This trend is shown in Fig.~\ref{fig:icews_triad_distribution} (a).

\paragraph{Structural balance theory:}
This theory predicts an evolution of the structure of these networks toward a state in which all violations of transitivity are eliminated. Thus, the theory focuses on the transitions of triads as states. In this theory, triads ---subsets of three nodes--- are considered the building blocks of relationships. Every possible triad of three countries among the core countries (134 countries) involves six edges in which each edge is either positive, negative or null. Classic balance theory assumes the absence of null edges, in which case there are 16 possible type of triads~\cite{sorensen1976stochastic, rawlings2017structural, nc2019, marvel2011}. In contrast, when null arcs are allowed, there are 138 possible types of triads some of which may entail one or more violations of transitivity. Ergo, this generalized structural balance theory is defined on sparse networks which are extremely realistic. The underlying idea for this generalization is that at any given time, if an edge has not been existed until now, it means there exists no social tension between those two nodes and that edge should not be considered part of a violation by classical axioms (See Fig. S8 in appendix showing total 138 possible triads).
To follow a tremendous literature on structural balance theory, we extend it on sparse networks using three existing models classical~\cite{cartwright1956structural}, clustering~\cite{davis1967clustering}, and transitivity~\cite{holland1971transitivity} (from least to most general respectively). Each model permits different triad set to be considered structurally balanced. The extension of these models are formally defined in Materials and Methods (see Table~\ref{tbl:balance_theory_definitions}).

Remarkably, we find that 91\% of the 392,084 observed triads in 23+ years are concentrated on only 10 triad types (out of 138). Fig.~\ref{fig:icews_triad_distribution} (d) displays this operative set of triad types. The types 6, 8 and 9 include one or more violations of transitivity, and the others do not. Fig.~\ref{fig:icews_triad_distribution} (c) shows the average distribution of operative triads over time and Fig.~\ref{fig:icews_triad_distribution} (b) depicts the temporal trajectory of the percentage of transitive triads. It is evident that structural balance does not always increase. Our finding based on Fig.~\ref{fig:icews_triad_distribution} (b) is that after a decrease of structural balance during the first periods, the network's trend is toward greater balance since 2006 onward. The main basis of the initial decline are conversions of null relations to positive relations and the associated proliferation of intransitive triad 9 configurations. Overtime, many of these violations of transitivity are then resolved by conversions to triad configurations that do not violate transitivity (triads 1, 4 or 10). Also, looking at distribution of triads (Fig.~\ref{fig:icews_triad_distribution} (c)), we see only about 8\% of triads (summation of volume of triads 6, 8 and 9) to be not-balanced over course of more than two decades. The evidence for this result is highlighted in our Markov chain analysis to which we now turn.

\subsubsection{Markov model on dynamic networks}
\begin{figure}[!ht]
    \centering
    \xincludegraphics[width=0.4\linewidth, label=\textbf{a)}]{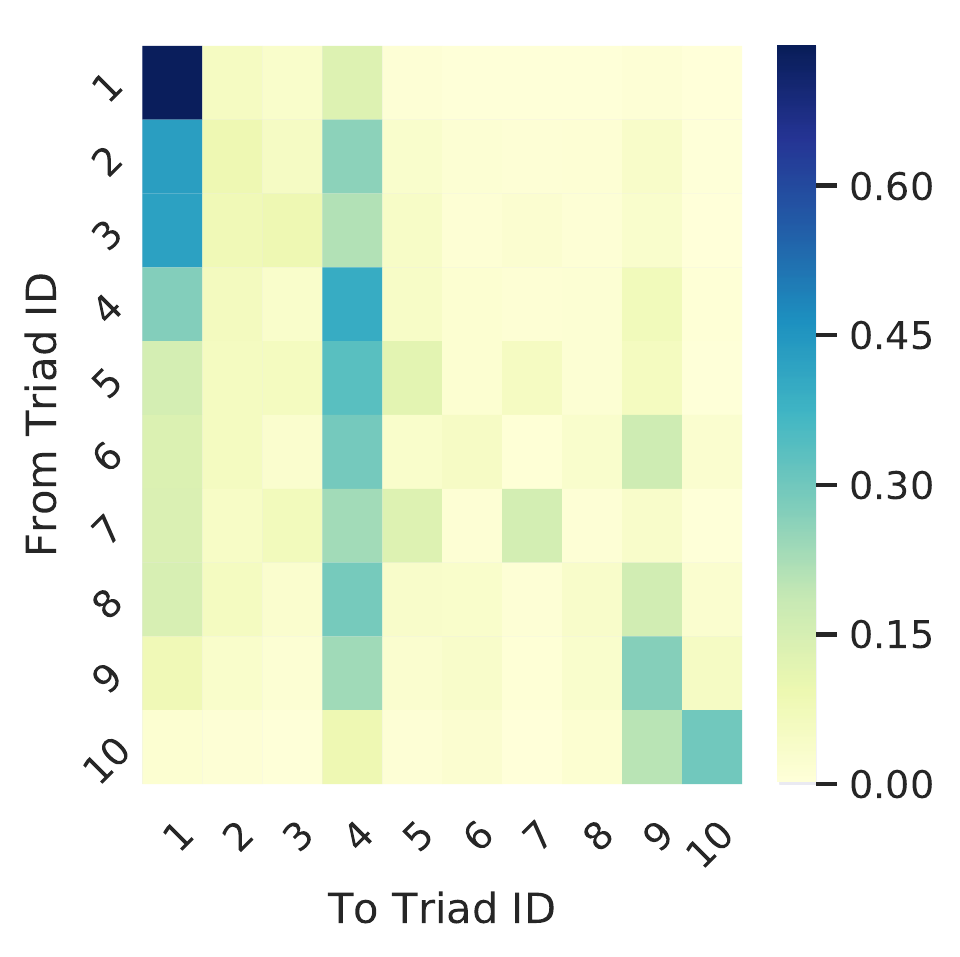}
    \xincludegraphics[width=0.4\linewidth, label=\textbf{b)}]{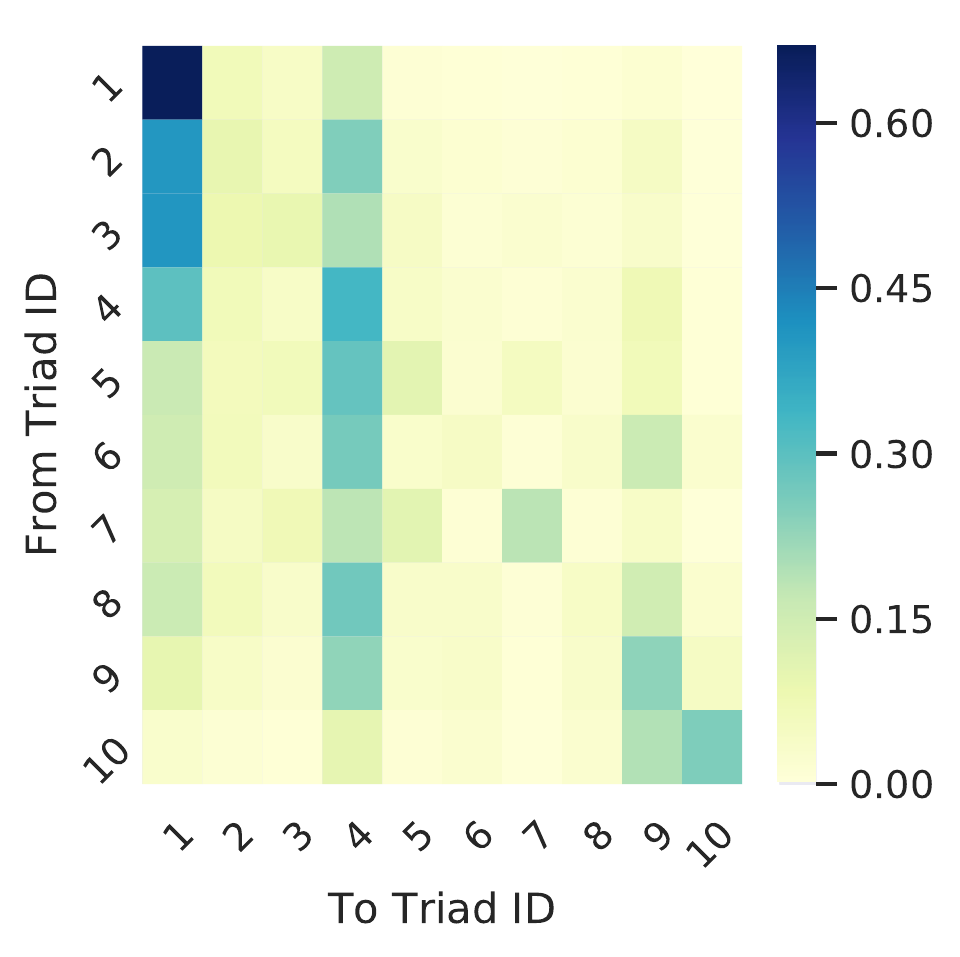}
    \xincludegraphics[width=0.4\linewidth, label=\textbf{c)}]{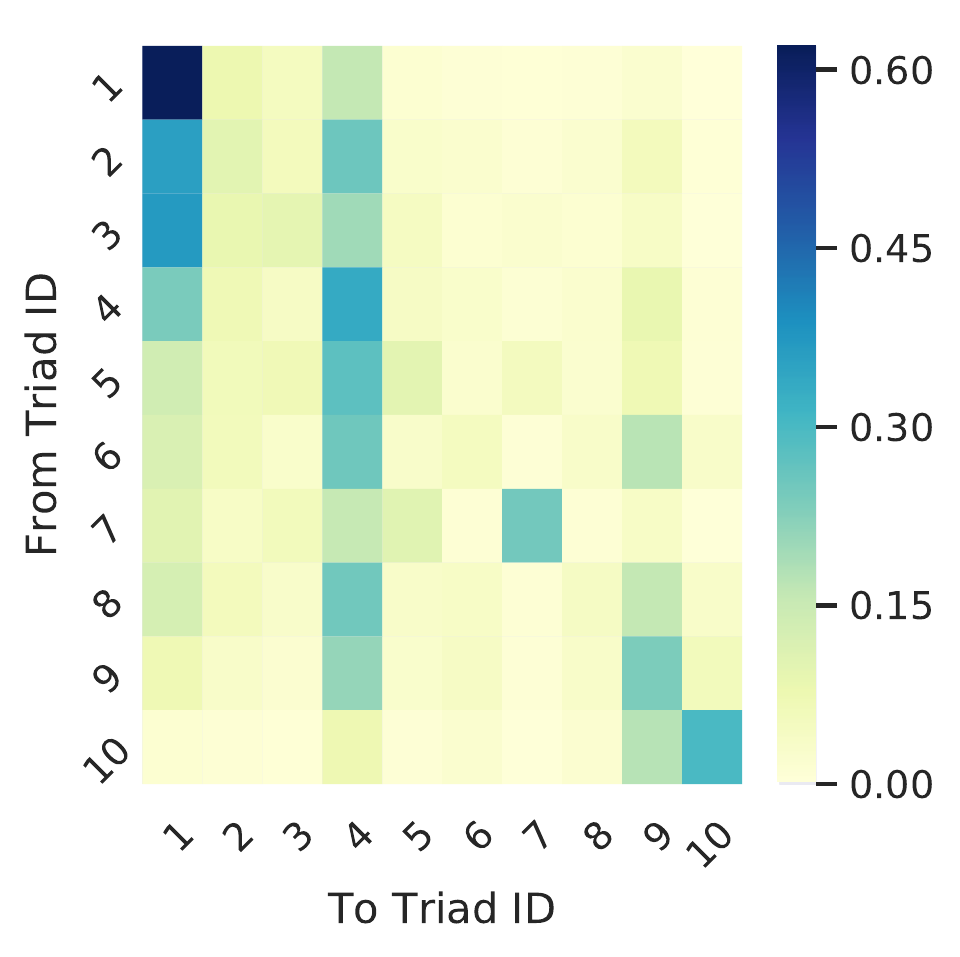}
    \xincludegraphics[width=0.4\linewidth, label=\textbf{d)}]{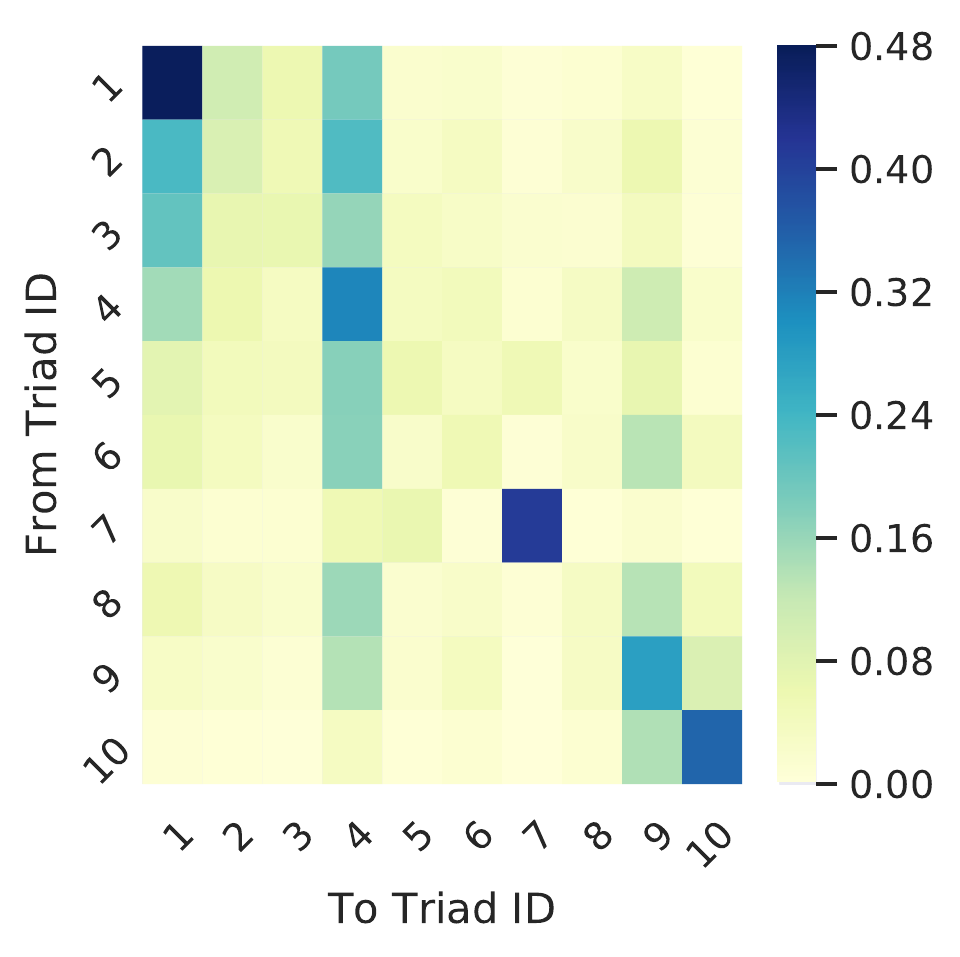}
    \xincludegraphics[width=0.4\linewidth, label=\textbf{e)}]{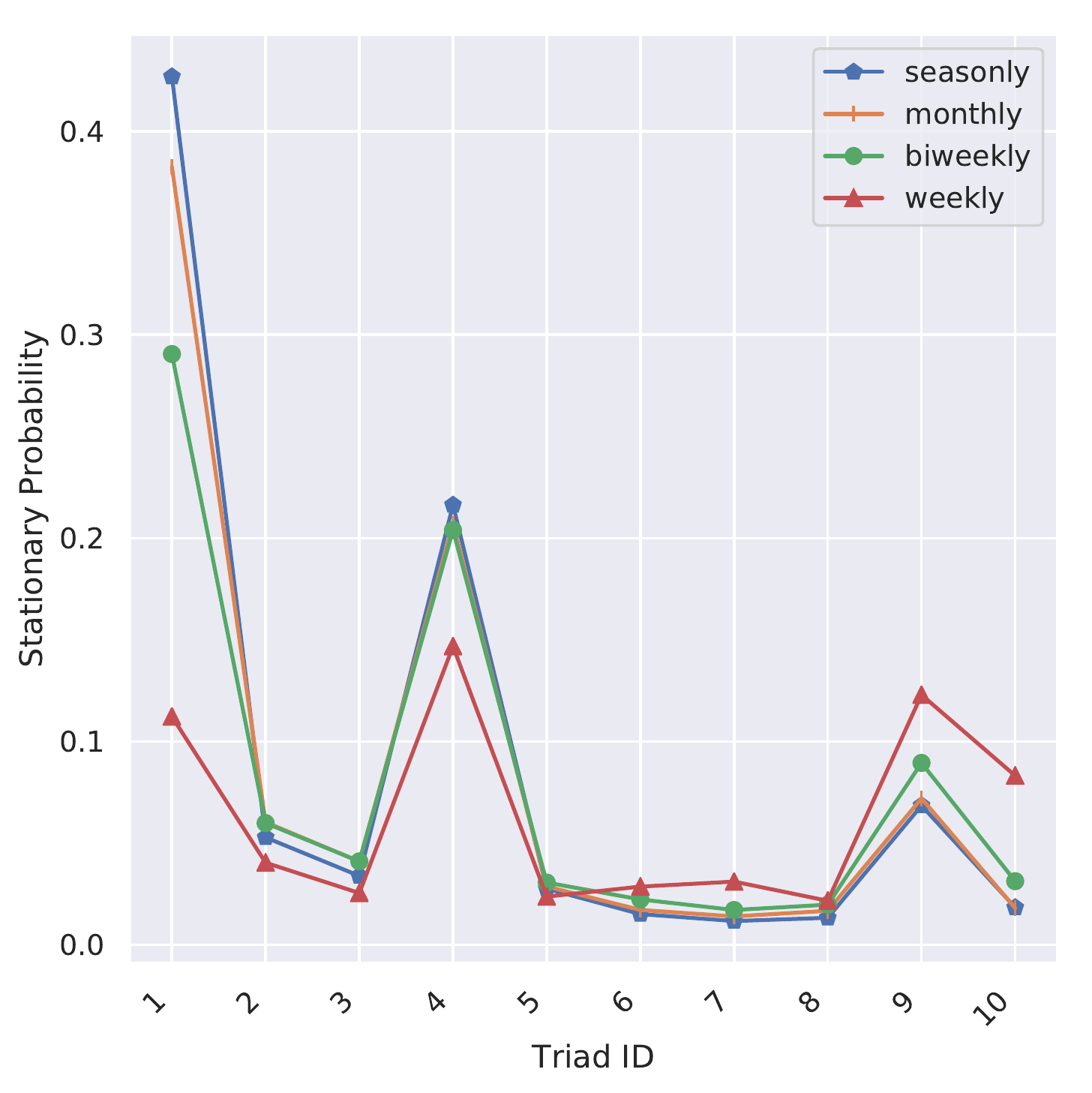}
    \caption{Average probability transition matrix in ICEWS dataset by aggregating dynamic networks of a) seasonally b) monthly c) biweekly d) weekly period. The transition matrix is row-stochastic and its elements falls into $(0, 1]$. There are 138 possible triads; only the sub-matrix of the 10 operative triads is shown (given in Fig.~\ref{fig:icews_triad_distribution} (e)). Sub-figure e) shows probability stationary distribution of the transition matrices, with the different aforementioned periods. The stationary distribution shows the state of the system under the condition that the Markov model persists. It appears that the probability transition matrix and stationary distribution are robust with respect to choice of period length.}
\label{fig:icews_average_transition_matrix}
\end{figure}

\hfill\\
Here, we present a Markov model of the dynamical system of the temporal transitions of the networks' triads that is not restricted to the operative set of triad types. This model provides a deeper image of the probabilistic micro-dynamics of the alterations of international appraisals during the 1995-2018 span of the available data. We compute the average probability transition matrix of the 138 possible triad types from networks aggregated over a three months period (seasonally). Interestingly, most of probabilities in this matrix are very close to zero and the dynamics of the system can be described only by a few states. For the sake of visualization, we can focus on the operative set of 10 triad types (described in Fig.~\ref{fig:icews_triad_distribution} (e)) and we show their transition probabilities in Fig.~\ref{fig:icews_average_transition_matrix} (a). The probability transition matrix is robust with respect to the choice of period. In Fig.~\ref{fig:icews_average_transition_matrix}, each sub-figure shows the transition matrix for period lengths: (a) seasonally (b) monthly (c) biweekly (d) weekly. The probability transition matrices look very similar. Quantitatively, the Pearson correlation between the flatten format of transition matrix in (a) with (b), (a) with (c), and (a) with (d) is 0.99, 0.98, and 0.86, respectively where all are statistically significant ($p <$ 0.05). Based on these transition matrices, one can see the triads 1, 4, 9 and 10 have large self-transition probabilities (high probability of transitioning from 1 to 1) and, thus, are most likely to persist. More precisely, Fig.~\ref{fig:icews_average_transition_matrix} (e) shows the stationary distribution of the Markov process. The summation of balanced triads in the stationary distribution is larger than 0.85. It appears that regardless of the definition of period, the Markov model predicts our empirical finding of a network evolution toward structural balance.

\begin{figure}[!ht]
    \centering
    \includegraphics[width=0.9\linewidth]{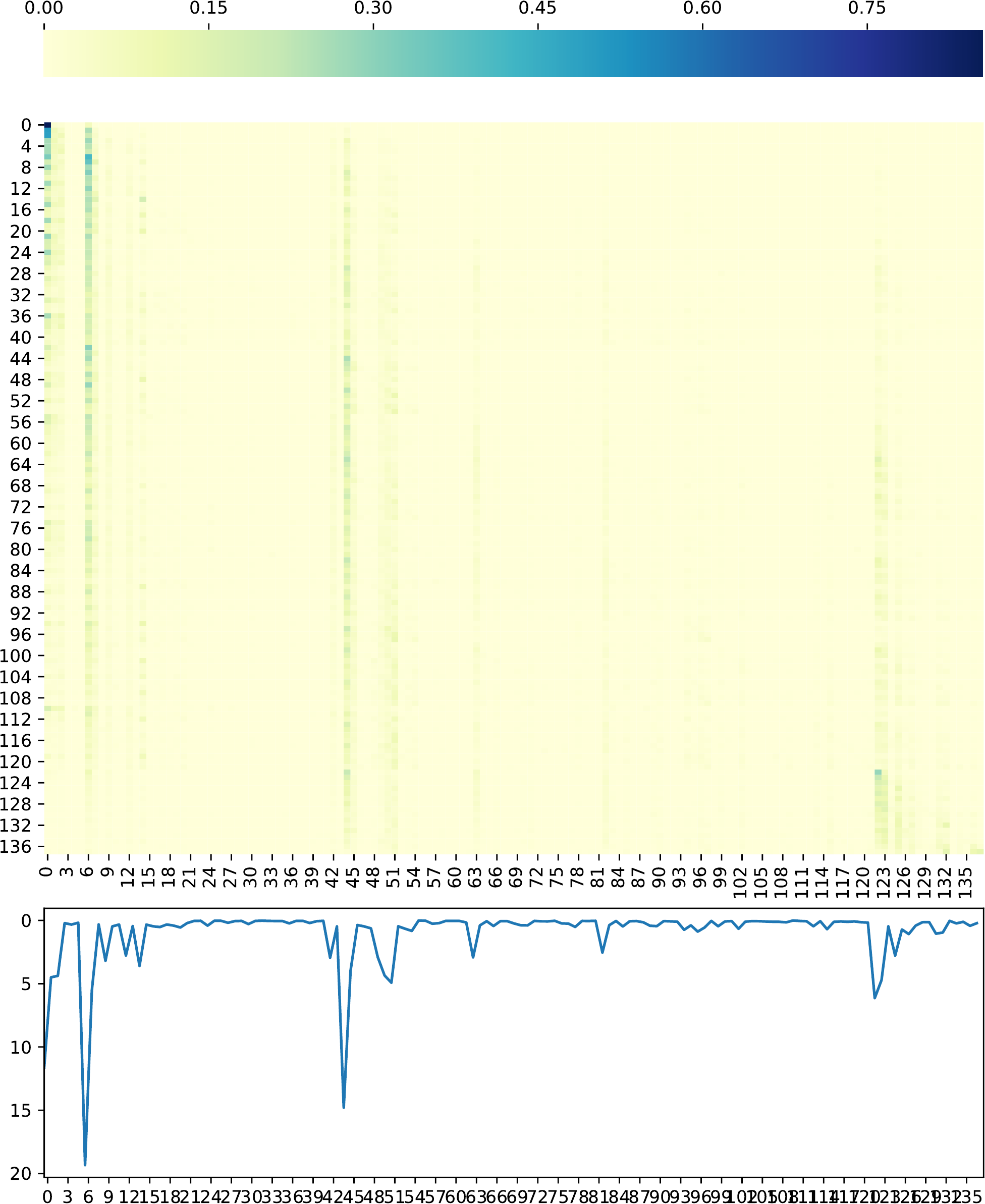} 
    \caption{Average probability transition matrix in ICEWS dataset from 101 transition matrices. States are 138 possible sparse triads. The transition matrix is row-stochastic and its elements falls into $(0, 1]$. The subfigure on the X-axis shows the summation of each column of the transition matrix. To have a better view, once can focus on ten aforementioned highly common triads' transition probability matrix in Fig.~\ref{fig:icews_average_transition_matrix}.}
\label{fig:icews_average_transition_matrix_all}
\end{figure}

\subsubsection{Time-varying Markov model on dynamic networks}\hfill\\

\begin{figure*}[!ht]
    \centering
    \includegraphics[width=1.05\linewidth]{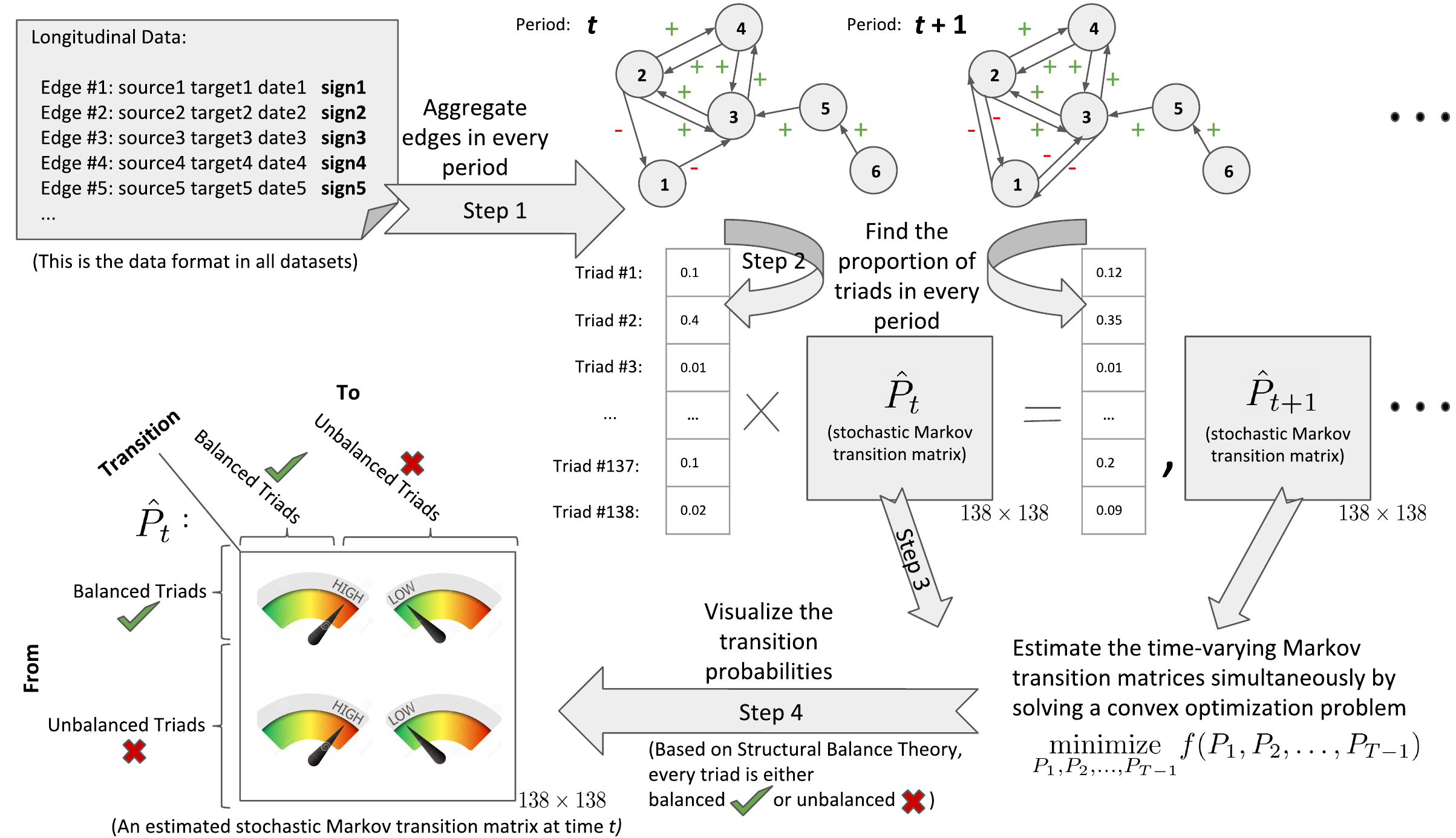}
    \caption{Estimation of transition probability matrices between balanced and unbalanced triads over a sequence of time periods. This figure illustrates the preprocessing steps, the optimization problem, and the results. By allowing null ties, there are 138 types of triads. $P_t$ represents the unknown transition matrix at time $t$ and $\hat{P_t}$ representing the estimated transition matrix at time $t$ (using ~\eqref{eq:objective_function}). $P_t$ is the variable of the optimization problem, and $\hat{P_t}$ is the solution. The four quadrants of $\hat{P_t}$ show the average estimated transition probabilities (see section Materials and 
    Methods for details). The result of this experiment is shown in Fig.~\ref{fig:boxplot_transition_probabilities}.}
\label{fig:method_description}
\end{figure*}

\begin{figure}[!ht]
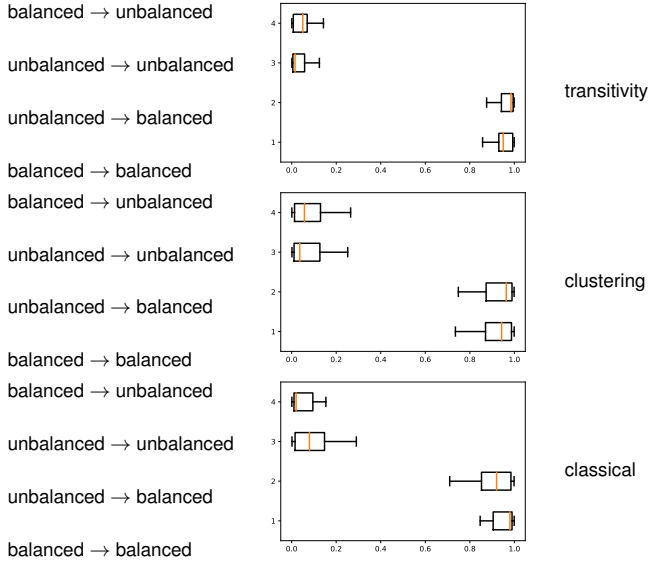

\begin{center}
\begin{tabular}{lllr}
\begin{tabular}[c]{@{}l@{}}balanced $\rightarrow$ unbalanced\\\\ unbalanced $\rightarrow$ unbalanced\\\\ unbalanced $\rightarrow$ balanced\\\\ balanced $\rightarrow$ balanced\end{tabular} & \adjustimage{width=0.2\textwidth, valign=m}{ICEWS_transitivity_nolabels} & transitivity \\

\begin{tabular}[c]{@{}l@{}}balanced $\rightarrow$ unbalanced\\\\ unbalanced $\rightarrow$ unbalanced\\\\ unbalanced $\rightarrow$ balanced\\\\ balanced $\rightarrow$ balanced\end{tabular} & \adjustimage{width=0.2\textwidth, valign=m}{ICEWS_clustering_nolabels} & clustering \\

\begin{tabular}[c]{@{}l@{}}balanced $\rightarrow$ unbalanced\\\\ unbalanced $\rightarrow$ unbalanced\\\\ unbalanced $\rightarrow$ balanced\\\\ balanced $\rightarrow$ balanced\end{tabular} & \adjustimage{width=0.2\textwidth, valign=m}{ICEWS_classical_nolabels} & classical
\end{tabular}
\end{center}
\caption{Estimated transition probability for classical-, clustering-, and transitivity-balanced and -unbalanced triads (the pipeline is described in Fig.~\ref{fig:method_description}). The x-axis shows the estimated probability (computed by solving ~\eqref{eq:objective_function}), the box is the interquartile range of the probability distribution, the orange line is the median of the distribution, and the whisker shows minimum and maximum of the range of the distribution. This figure shows that the probability of transitions from unbalanced triads to balanced ones is significantly higher than the opposite transitions. Also, the probability of remaining balanced is more likely than the probability of remaining unbalanced. These findings hold regardless of the definition of balance (see section Generalized Structural Balance). Our experiments show that movement toward balance is not limited to a dataset or type of data.}
\label{fig:boxplot_transition_probabilities}
\end{figure}

Our results show that the probability of transitions to and remaining in balanced states are statistically significant in every period over the 23+ year data span. We find the same results with a novel time-varying Markov model in which the transition matrix can smoothly change and is learned via a convex optimization scheme (see section Materials and Methods). Fig.~\ref{fig:method_description} describes the analysis pipeline of this model. Our Fig.~\ref{fig:boxplot_transition_probabilities} results further support the conclusion that structural balance drives the dynamics of the system. All unbalanced triads have an estimated high mean probability and small standard deviation on transitions into balanced triads, and the balanced triads have an estimated high mean probability and small standard deviation of remaining balanced. Note the distinctive separation of the transition probabilities.

Applying this experiment on other datasets we posit this result not only holds in our focal network of international relations, but also in two longitudinal datasets on financial bitcoin trust networks~\cite{kumar2016edge, kumar2018rev2}. Due to focus of this paper and lack of space we do not include those figures. Results show although international and financial networks are very different, our findings on transition toward balance generalize across all three datasets. Therefore, it appears that transitions toward balance are ubiquitous (i) regardless of the definition of balance, (ii) regardless of the setting (international news networks or financial networks), and (iii) regardless of the type of actor (individuals or countries).

\subsection{Diminishing Variability among Nations}\hfill\\
\subsubsection{Qualitative Relation with Exogenous Shocks}\hfill\\
The transition matrices are stable over time, as measured by the Frobenius norm difference of consecutive matrices (Fig.~\ref{fig:frobenius_vs_trade}). Interestingly, the Frobenius norm of transition matrices declines smoothly over time. We call this phenomenon, the stability of the dynamics. This finding is aligned with the fact that the number of wars per pair of countries in the past 50 years was roughly a 10th as high as it was from 1820 to 1949. Fig.~\ref{fig:frobenius_vs_trade} (a) suggests that the disruptions to this trend are associated with important shocks such as the September 11th, 2001 attacks (9/11).

\subsubsection{Quantitative Relation with International Trade Activity}
\begin{figure}[!ht]
\centering
\xincludegraphics[width=1.2\linewidth, label=\textbf{a)}]{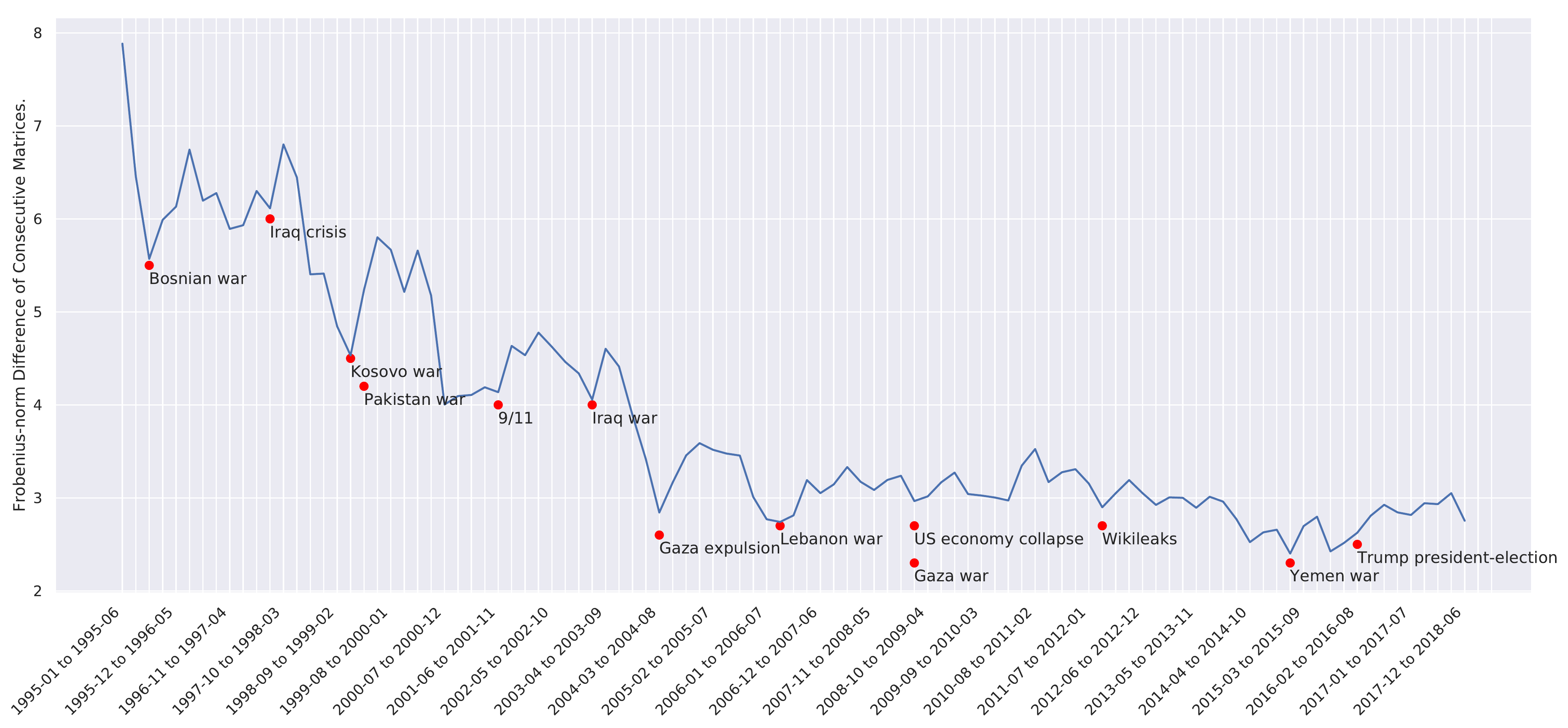}
\xincludegraphics[width=0.8\linewidth, label=\textbf{b)}]{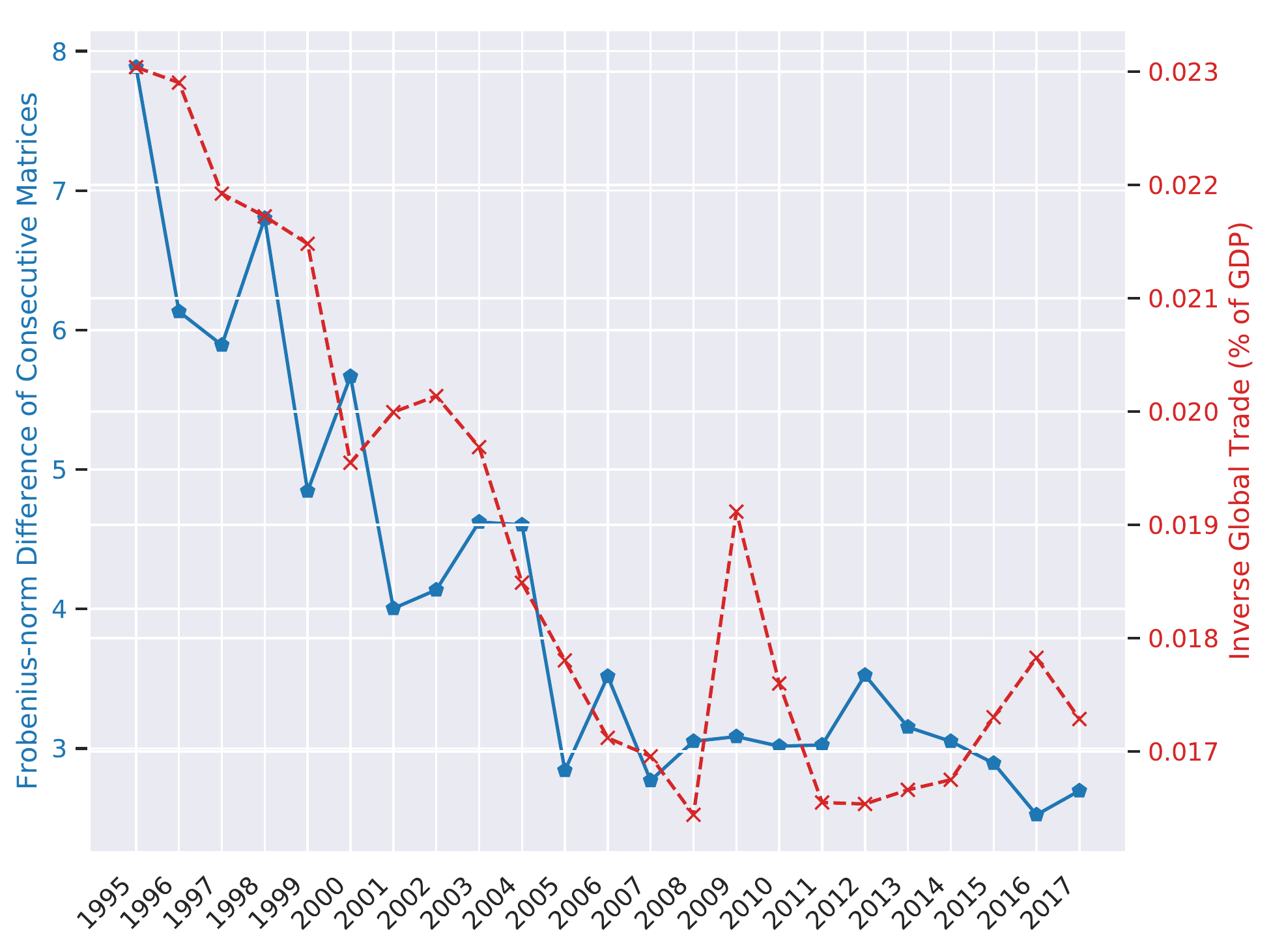}
\caption{a) Frobenius norm difference of consecutive empirical transition matrices. The figure shows the Markov chain of triads becomes stable over time as the Frobenius norm of difference decreases over time. Notably, we find most global exogenous events such as wars (many mentioned on the figure) trigger disturbances in the dynamics of relationships. b) For more quantitative result, the same Frobenius norm difference of consecutive empirical transition matrices (in straight blue line), versus international trades (in dashed red line). The figure shows a reverse relation between change in the system and trades, meaning as relationships among countries has became stable, the trades among which, have been increased. These two are statistically correlated (Pearson correlation coefficient of 0.88 ($p < $ 1e-07)) and amount of trades statistically Granger causes~\cite{granger1988causality} the relationship among nations ($p < $ 1e-03). Aligned with this finding, earlier research~\cite{jackson2015networks, barbieri2009trading, martin2008make, oneal1999assessing, hegre2010trade} also claimed increased trades have decreased countries' incentive to attack each other, leading to a stable and network of alliances.}
\label{fig:frobenius_vs_trade}
\end{figure}

\hfill\\
Additionally, inspired by previous research~\cite{jackson2015networks, barbieri2009trading, martin2008make, oneal1999assessing, hegre2010trade}, using the data on international trades among nations since 1995, in Fig.~\ref{fig:frobenius_vs_trade} (b), we find a statistically significant correlation between the Frobenius norm difference of consecutive matrices and inverse of global trades. World trade data shows the global trades among all countries in world as of the percentage of each countries' GDP, which is extracted from The World Bank national accounts data (see section Data Availability for details). The stability of dynamics and global trades are correlated in past 23+ years (Pearson correlation coefficient of 0.88 ($p <$ 1e-07)). Fig.~\ref{fig:frobenius_vs_trade} shows that as relationships among countries have become stable over the years, the volume of trades has been increased.

Moreover, our causality test shows evidently the more global trades there are, the more stable the relational dynamics become and vice versa over the course of two decades. Granger causality test~\cite{granger1988causality} shows an statistically significant effect of global trades on the stability of the dynamics ($p <$ 1e-03), and also shows a feedback effect ($p <$ 1e-02). The causality tests are found to be statistically significant using both F-test and chi2-test with $\#\text{lags} = 1$. 

This result simply means as the relations among countries become more stable, they are more willing to trade for economical benefits, and on the other hand when they internationally trade with one another, they are willing to have a stable relationship with one another. In this study, stability is captured with Markov transition matrices of triads over the course of two decades. This finding comports with the seminal study by Jackson \emph{et al.}~\cite{jackson2015networks} and its finding on stabilizing the international conflicts by the decrease of the number of wars among nations as the same rate as the increase in global trades.

\section{Discussion}
Balance theory has triggered a literature of efforts to specify the mechanisms that alter interpersonal appraisal networks~\cite{heider1946attitudes, cartwright1956structural, friedkin2011formal} towards states of structural balance. This theory is also associated with research on international relations. However, despite the need for longitudinal data to recover the underlying dynamics of balance theory, such investigations have been rare. We have leveraged an extensive longitudinal dataset to advance the  research on the evolution of the network of global international relations, and the basic science on balance theory. We find consistently high probabilities of transition toward and remaining in balanced triads and not vice versa. We believe that balance theory's prediction of a structural evolution toward balanced states is sound. Also, we find that the network dynamics of international relations over the past 23+ years have been toward structural stability, consistent with balance theory expectations, with occasional shocks of large scale international events on the trajectory of the global network.


\section{Materials and Methods}\label{sec:matmethods}
\subsection{Definitions}\hfill\\
In Table~\ref{tbl:definitions}, we summarize the major notations used throughout the present section.

\begin{table}[!ht]
\centering
\caption{Description of basic symbols.}
\begin{tabular}{ll}
Symbols & Definition \\
\midrule
$V$ & The set of nodes in a network\\
$E$ & The set of edges in a network\\
$e_{ij}$ & Directed edge from node $i$ to node $j$ \\
$A_t$ & Adjacency matrix of directed and signed network at time $t$ \\
$P_t$ & Markov transition probability matrix from time $t$ to $t+1$ \\
$\hat{P_t}$ & Empirical Markov transition probability matrix from time $t$ to $t+1$ \\
$\widetilde{P_t}$ & Estimated time-varying Markov transition probability matrix \\
$T$ & The number of available time periods \\
\bottomrule
\end{tabular}
\label{tbl:definitions}
\end{table}

\subsection{Generalized Structural Balance}\label{sec:generalized_sbt}\hfill\\
In order to give formal definitions of different versions of generalized balance theory, we start with Heider's~\cite{heider1946attitudes, heider2013psychology} four axioms: 
  
\begin{itemize}[noitemsep,nolistsep]
    \item \textbf{A1}- Friend of a friend is a friend.
    \item \textbf{A2}- Friend of an enemy is an enemy.
    \item \textbf{A3}- Enemy of a friend is an enemy.
    \item \textbf{A4}- Enemy of an enemy is a friend.
\end{itemize}

\noindent
Classical balance theory assumes a fully connected network~\cite{heider1946attitudes, davis1967clustering, johnsen1985network, rawlings2017structural}.
We generalize three definitions of balance, based on the above axioms, to networks with null edges.
The value of $e_{ij}$ can have negative, positive or zero (null) value. Out of 138 possible triads, 93 are transitive-balanced (67\%), 44 are cluster-balanced (32\%), and 24 are classical-balanced (17\%). Remarkably, we find a large set of forbidden triad types are transitioning to a relatively small set of permissible triad types (Fig.~\ref{fig:boxplot_transition_probabilities}).
In balance theory literature, the concept of sparse balance theory has been before addressed as incomplete awareness~\cite{montgomery2009balance, de1999sign}. The concept has been motivated by the empirical evidence that affective relations are signed but seldom complete --- actors may be neutral toward each other or there may be null or unobserved edges. Cartwright and Harary~\cite{cartwright1956structural} define balanced cycles on networks with missing edges such that the only condition is as cycles containing an even number of negative edges. In this work, we extend the analysis of networks with null edges to the general case of sparse triads.

Assume every directed edge $e$ has value $\in \{-1, 0, +1\}$. For every three distinct nodes $i, j, k$, to be considered as a balanced triad, the following condition, for any permutation of the nodes, needs to hold

\begin{table}[!ht]
\caption{Different definitions of sparse structural balance theory that generalize existing definitions of balance. Transitivity is the most general model that only requires the first axiom.}
\begin{tabular}{|l|l|c|}
\hline
Balance Model & Heider Axioms & Structural Equation (condition) \\ \hline
classical~\cite{cartwright1956structural} & A1, A2, A3, A4 & {$\!
    \begin{aligned}
        \forall&  i, j, k \in V, \text{\; for every combination:}\\
        & \text{if}\; e_{i k} \neq 0\; \text{and} \; e_{k j} \neq 0\\
        & \text{then}\;\; e_{i j} = e_{i k} e_{k j} \text{\;\;should be valid}
    \end{aligned}$} \\ \hline
clustering~\cite{davis1967clustering} & A1, A2, A3 & {$\!
    \begin{aligned}
        \forall&  i, j, k \in V, \text{\;for every combination:}\\
        & \text{if}\; e_{i k} \neq 0\; \text{and}\; e_{k j} \neq 0\\
        & \text{and}\;\; (e_{i k} > 0\; \text{or}\; e_{k j} > 0)\\
        & \text{then}\;\; e_{i j} = e_{i k}e_{k j} \text{\;\;should be valid}
    \end{aligned}$}  \\ \hline
transitivity~\cite{holland1971transitivity} & A1 & {$\!
    \begin{aligned}
        \forall&  i, j, k \in V, \text{\;for every combination:}\\
        & \text{and}\;\; e_{i k} > 0\; \text{and}\; e_{k j} > 0\\
        & \text{then}\;\; e_{i j} = e_{i k}e_{k j} \text{\;\;should be valid}
    \end{aligned}$} \\ \hline
\end{tabular}
\label{tbl:balance_theory_definitions}
\end{table}

\subsection{Network Extraction}\hfill\\
Networks are extracted by aggregating edges in predetermined periods. If the period length is too short we would not obtain sufficient information, while too long periods would decrease the granularity of the observations. We use 12 weeks ($\sim$1 quarter) as the period duration. Note, Fig.~\ref{fig:icews_average_transition_matrix} shows results based on transition matrices are robust with respect to the choice of period length. Consequently, for ICEWS dataset, we have 103 networks. For a given network, the appraisal between nodes $i$ and $j$ is determined by the sign of summed edge weights of all directed edges observed between them (edge $(i, j)$), during that time period, as described by

\begin{equation}\label{eq:edge_sign}
    A_{ij}(t_k) = \operatorname{sign}\Bigg( \sum_{t_k \leq t \leq t_{k+1}} w_{ij}(t) \Bigg),
\end{equation}
\noindent
where $A(t_k)$ shows the adjacency matrix at period $t_k$.

\subsection{Empirical Markov Transition Matrices}\hfill\\
For each consecutive observation period $(t,\;t+1)$, we compute $P_{ij}(t)$, the number of triads of type $i$ that moved to type $j$ from period $t$ to $t+1$. In fact, for every three nodes in the network, we find the corresponding triad type, at time $t$ and time $t+1$, say triad type $i$ and triad type $j$, respectively. Then increment the number of transitions happening from $P_{ij}(t) \leftarrow P_{ij}(t) + 1$. Thus, row $i$ sums to $P_{i \star}(t)$, the number of triads of type $i$ at time $t$, while $P_{\star j}(t+1)$  is the number of triads that have transitioned to type $j$ at time $t+1$. Using the transition matrix $P_{ij}(t+1)$, the transition probabilities can be estimated to obtain the transition probability matrix. These quantities can be arranged in a matrix and normalized by the sum of every row. Therefore, we have row-stochastic transition matrix $P$ where each $P_{ij}(t)$ is conditional on $i$ only, and not on prior states occupied by the triad. By the Markov property, they are identical for all triads, and they converge to a stationary probability distribution.

\subsection{Estimating Time-varying Markov Transition Matrices}\hfill\\
Estimating Markov transition matrices via counting the observed transitions only takes into account the subsequent periods and therefore does not take advantage of any other similarity among transition matrices. The goal is to have a method that while keeping the Markov attribute of the system, allows for transition matrices to transfer information based on the existing assumptions in the literature, such as applying smoothness to estimate a more accurate set of transition probability matrices. Hence, we use time-varying Markov chains to capture the most out of the observed transitions in the data. The length of the longitudinal data, in this study, entails having statistical sufficiency to apply a nonparametric convex method to accurately estimate the transition probability matrices directly from the data.

\subsubsection{Model Formulation}\hfill\\
For a network $A_t$ at time $t$, we count the occurrences of each of three nodes, and classify each into one of 138 possible triads. There are $T + 1$ periods and thereupon $T + 1$ networks.

There are $m$ entities (triads in a dynamic network), that in parallel, change states for $T + 1$ periods of time. As discussed before, there are 138 triad types ($l_1, \dots, l_{138}$). Each empirical Markov probability transition matrix, $\hat P_t$, is computed as follows

\begin{equation}
    \hat P_t = \frac{\sum\limits_{i=1}^{m} \indicator(S_t^{(i)}=l_t,\; S_{t-1}^{(i)}=l_{t-1})}{\sum\limits_{i=1}^{m} \indicator(S_{t-1}^{(i)}=l_{t-1})}
\end{equation}

\noindent
where each $\hat P_t$ is a $138 \times 138$ matrix and there exist $\hat P_1, \dots, \hat P_T$ as empirical transition matrices. Fig.~\ref{fig:icews_average_transition_matrix} shows the average empirical Markov transition matrices for different choice of period length. Now, we formalize an optimization problem to account for potential error in each empirical transition matrix as being optimized to be close to the true underlying time-varying Markov transition matrix as
\begin{equation}
    \mathbf{P_t} \sim \mathbf{\hat{P_t}}.
\end{equation}

The algorithm considers these empirical transition matrices as the input, and estimates all latent transition matrices, simultaneously. By definition, a Markov transition matrix, $\mathbf{P_t}: 138 \times 138$, should be ergodic, aperiodic and irreducible. Simply put, in the considered Markov chain, it should be possible to be in any state and also should be possible to get to any state from any state. Thus, there are $T$ probability Markov transition probability matrices and every $\mathbf{P_t}$ needs to satisfy

\begin{equation}\label{eq:transition_matrices}
    0 < (\mathbf{P_t})_{ij} \leq 1, \quad \forall i, j \in \{1,\dots,138\} \text{ and } t \in \{1,\dots,T\}.
\end{equation}

By definition, Markov transition probability matrices should be row-stochastic --- every row is sum up to 1. That is,
\begin{equation}\label{eq:transition_matrix_condition}
    \ones^T \mathbf{P_t} = \ones^T.
\end{equation}

Regarding the objective function, based on previous studies dealing with time-varying Markov chains~\cite{chiba2017time}, we make an assumption that subsequent transition matrices are similar to each other; the changes happen smoothly.
\begin{equation}
    \mathbf{P_t} \sim \mathbf{P_{t-1}}.
\end{equation}

\subsubsection{Optimization Problem}\hfill\\
To estimate all unknown transition matrices simultaneously ($T$ matrices of size $138 \times 138$), we setup an appropriate optimization problem; we shall solve the optimization problem using convex optimization methods.

\begin{equation}\label{eq:objective_function}
    \begin{array}{lll}
        & \widetilde{\mathbf{P_1}}, \dots, \widetilde{\mathbf{P_T}} = \\ 
        & \minimize\limits_{\mathbf{P_1}, \dots, \mathbf{P_T}} & \halft \sumtot \left\lVert\hat{\mathbf{P_t}} - \mathbf{P_t} \right\rVert_F^2 + \sumtwotot \psi\bigg(\mathbf{P_t} - \mathbf{P_{t-1}}\bigg), \\
        & \mbox{where} & \psi(X) = \lambda_1 \|X\|_1 + \lambda_2 \|X\|_2,\\\\
        & \subjectto & (\mathbf{P_t})_{ij} > 0,\; \forall i, j \in [1, 138], \\
        & & (\mathbf{P_t})_{ij} \leq 1,\; \forall i, j \in [1, 138], \\
        &           & \ones^T \mathbf{P_t} = \ones^T,
    \end{array}
\end{equation}

\noindent
where $\widetilde{P_t}: t = 1, \dots, T$ are the estimated transition matrices and their results are shown in Fig.~\ref{fig:boxplot_transition_probabilities}. In ~\eqref{eq:objective_function}, we apply two forms of smoothness on subsequent Markov transition matrices: $l2$-norm (Frobenius norm), also called group lasso that enforces a small amount of change in transition matrices, and $l1$-norm, also called fused lasso that induces a sparse solution with respect to the changes in matrices. In optimization literature, this criterion is called sparse group lasso~\cite{friedman2010note}. Together they encourage subsequent Markov transition matrices to only deviate from each other with small values and in only a few cells.

This formulation allows for learning a separate model for each transition between time periods while inferring information globally across all periods. Non-parametric estimating all transition matrices together via an optimization problem decreases the chance of overfitting. This method also allows for finer time windows than otherwise, and provides a better inference granularity in time. Consequently, even with few observations of the data, we end up with an accurate estimation for time-varying transition matrices. Algorithm~\ref{alg:markov_chain_estimation} illustrates the steps for estimating the time-varying Markov chains.

The problem in ~\eqref{eq:objective_function} is convex. The reason is that the objective function is a summation of two norms which are convex, all of the inequality constraints are convex, and all equality constraints are affine. Therefore, the problem is convex~\cite{boyd2004convex}, it has a globally optimal solution, and we solve this equation by a convex optimization solver, CVXPY~\cite{cvxpy}.

Note, in appendix, we provide a novel proof for optimality of the convergence rate given a bound in probability for the proposed optimization model of simultaneous leaning of all transition matrices in ~\eqref{eq:objective_function}.

\begin{algorithm}[!ht]
    \SetAlgoLined
    \textbf{Input:} Signed directed networks over time $\{A\}_{t=1}^T$\;\\   
    \textbf{Output:} Estimated transition matrices $\{\widetilde{P}\}_{t=1}^{T-1}$\;\\\\  
    Tune the hyper-parameters $\lambda_1$, and $\lambda_2$\;\\
    \For{\text{time starting from} $t=2$ \text{to} $T-1$}
    {
        $\hat{\mathbf{P_t}} := \text{zero matrix}[138, 138]$\;\\
        \For{\text{every triplet of nodes} $i, j, k$ \text{in set of nodes} $V$}
        {
          $\text{triad1} := \text{type}(A_{t-1}(i, j, k))$\;\\
          $\text{triad2} := \text{type}(A_{t}(i, j, k))$\;\\
          $\hat{\mathbf{P_t}}[\text{triad1}, \text{triad2}] := \hat{\mathbf{P_t}}[\text{triad1}, \text{triad2}] + 1$\;
        }
        $\hat{\mathbf{P_t}} := \hat{\mathbf{P_t}} / \big(\ones^T\hat{\mathbf{P_t}}\big)$\;
    }
    Solve the convex optimization problem in~\eqref{eq:objective_function}\;
    \caption{Non-parametric estimation of all time-varying transition matrices simultaneously using a convex objective function.}
\label{alg:markov_chain_estimation}
\end{algorithm}

\subsubsection{Model Comparison}
\begin{figure}[!ht]
    \centering
    \includegraphics[width=0.9\linewidth]{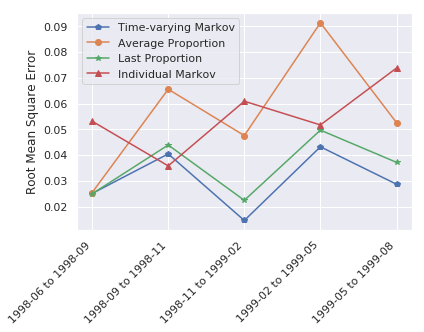}
    \caption{Root Mean Square Error (RMSE) of different baselines and time-varying Markov chain models to predict unseen triad proportions in ICEWS dataset (the smaller the better). The proposed method is named {\em Time-varying Markov}. {\em Average Proportion} computes the average of each triad type proportion up to the current time. {\em Last Proportion} assumes the new proportion is exactly the last one, and due to the origin of being smooth, it is highly competitive. {\em Individual Markov} shows the method of counting each Markov models based on only consecutive times. The proposed method using time-varying Markov chain estimation surpassed the baselines' accuracy.}
    \label{fig:icews_rmse}
\end{figure}

\hfill\\
In order to test the predictability of the estimated transition matrices, we predict unseen proportion of the unseen triads using the proposed algorithm as compared to competitive baseline methods. For instance, assume we have $T$ periods. We hold out the proportion of triads at time $T$ and train on periods of 1 up to $T-1$ using Algorithm~\ref{alg:markov_chain_estimation}. Consequently, the estimated transition matrix for time $T-1$ to $T$ is multiplied with proportion at $T-1$ to give us the predicted proportion at $T$. We apply one step-ahead forecast, for multiple times, in each time retraining the model up to the last held out time. To this end, we have a more descriptive picture of the prediction power of the proposed method. As the forecast metric, we compute Root Mean Squared Error (RMSE) of the difference between the predicted proportion with the ground truth. Random prediction is neglected due to its significantly worse accuracy compared to other baselines. Consequently, we compare our forecast to a baseline of simply predicting the last time steps proportions, and versus an average of all preceding time steps' proportions.
Fig.~\ref{fig:icews_rmse} depicts that the proposed algorithm outperforms the baselines and makes accurate forecasts of the proportion of triads in the subsequent time period in the ICEWS dataset. We also use this forecasting method for multiple steps, and apply on a validation set of the periods, to carefully tune the hyper-parameters of algorithm~\ref{alg:markov_chain_estimation}. In this study, we set $\lambda_1 = 0.5$ and $\lambda_2 = 0.05$.

A future direction to improve the predictability of the model is to remove the Markov assumption in the modeling. We can let models find the best number of previous periods which should be taken into account for predicting the proportion of triads. Recursive neural networks can model the non-Markovian aspect inherent in the data.


\section*{Acknowledgement}
This work is supported by a UC Multicampus-National Lab Collaborative
Research and Training (UC-NL-CRT) grant, titled ``Political Conflict and Stability in Dynamic Networks,'' under grant number LFR-18-547591, and by the U.S. Army Research Laboratory and U.S. Army Research Office under grant number W911NF-15-1-0577.

We thank Prof. Yu-Xiang Wang for his instructive comments on the proposed optimization model and thank Haraldur Hallgrimsson for his comments on the manuscript.

\section*{Author Contributions}
O.A. analyzed the data and implemented the methods. O.A. and N.E.F. led the study. F.B. and A.K.S. actively participated in the discussion and gave critical comments. All authors substantially contributed to the design of the analysis and the writing of the paper.

\section*{Data Availability}\label{sec:data_availability}
The data is publicly available as the so-called ``Integrated Crisis Early Warning System (ICEWS) Dataverse'' by Boschee \emph{et al.}~\cite{boschee2015icews} under \url{https://doi.org/10.7910/DVN/28075}. The hyperlink also includes a document about the ICEWS event coding protocols.

The World Bank trade data (\% of GDP) is also publicly available under \url{https://data.worldbank.org/}.

\section*{Code Availability}
The source code is publicly available under \url{https://github.com/omid55/dynamic_sparse_balance_theory}.

\section*{Additional information}
\subsection*{Competing interests} The authors declare no conflicts of interest.

\subsection*{Reprints and permission} information is available online under \url{http://npg.nature.com/reprintsandpermissions/}.

\section*{References}
\bibliography{biblio}

\begin{thebibliography}{53}
\providecommand{\natexlab}[1]{#1}
\providecommand{\url}[1]{\texttt{#1}}
\expandafter\ifx\csname urlstyle\endcsname\relax
  \providecommand{\doi}[1]{doi: #1}\else
  \providecommand{\doi}{doi: \begingroup \urlstyle{rm}\Url}\fi

\bibitem[Jackson and Nei(2015)]{jackson2015networks}
Matthew~O Jackson and Stephen Nei.
\newblock Networks of military alliances, wars, and international trade.
\newblock \emph{Proceedings of the National Academy of Sciences}, 112\penalty0
  (50):\penalty0 15277--15284, 2015.

\bibitem[Zheng et~al.(2015)Zheng, Zeng, and Wang]{zheng2015social}
Xiaolong Zheng, Daniel Zeng, and Fei-Yue Wang.
\newblock Social balance in signed networks.
\newblock \emph{Information Systems Frontiers}, 17\penalty0 (5):\penalty0
  1077--1095, 2015.

\bibitem[McDonald and Rosecrance(1985)]{mcdonald1985alliance}
H~Brooke McDonald and Richard Rosecrance.
\newblock Alliance and structural balance in the international system: A
  reinterpretation.
\newblock \emph{Journal of Conflict Resolution}, 29\penalty0 (1):\penalty0
  57--82, 1985.

\bibitem[Easley and Kleinberg(2010)]{easley2012networks}
D.~Easley and J.~Kleinberg.
\newblock \emph{Networks, Crowds, and Markets: Reasoning About a Highly
  Connected World}.
\newblock Cambridge University Press, 2010.
\newblock ISBN 0521195330.

\bibitem[Harary(1961)]{harary1961structural}
Frank Harary.
\newblock A structural analysis of the situation in the {Middle East} in 1956.
\newblock \emph{Journal of Conflict Resolution}, 5\penalty0 (2):\penalty0
  167--178, 1961.

\bibitem[Doreian and Mrvar(2015)]{doreian2015structural}
Patrick Doreian and Andrej Mrvar.
\newblock Structural balance and signed international relations.
\newblock \emph{Journal of Social Structure}, 16:\penalty0 1, 2015.

\bibitem[Belaza et~al.(2017)Belaza, Hoefman, Ryckebusch, Bramson, van~den
  Heuvel, and Schoors]{belaza2017statistical}
Andres~M Belaza, Kevin Hoefman, Jan Ryckebusch, Aaron Bramson, Milan van~den
  Heuvel, and Koen Schoors.
\newblock Statistical physics of balance theory.
\newblock \emph{PLOS One}, 12\penalty0 (8):\penalty0 e0183696, 2017.

\bibitem[Heider(1946)]{heider1946attitudes}
Fritz Heider.
\newblock Attitudes and cognitive organization.
\newblock \emph{The Journal of Psychology}, 21\penalty0 (1):\penalty0 107--112,
  1946.

\bibitem[Cartwright and Harary(1956)]{cartwright1956structural}
Dorwin Cartwright and Frank Harary.
\newblock Structural balance: {A} generalization of {H}eider's theory.
\newblock \emph{Psychological Review}, 63\penalty0 (5):\penalty0 277, 1956.

\bibitem[Gellman(1989)]{gellman1989elusive}
Peter Gellman.
\newblock The elusive explanation: balance of power ‘theory’ and the
  origins of {World War I}.
\newblock \emph{Review of International Studies}, 15\penalty0 (2):\penalty0
  155--182, 1989.

\bibitem[Antal et~al.(2006)Antal, Krapivsky, and Redner]{antal2006social}
Tibor Antal, Paul~L Krapivsky, and Sidney Redner.
\newblock Social balance on networks: The dynamics of friendship and enmity.
\newblock \emph{Physica D: Nonlinear Phenomena}, 224\penalty0 (1-2):\penalty0
  130--136, 2006.

\bibitem[Moore(1978)]{moore1978international}
Michael Moore.
\newblock An international application of {H}eider's balance theory.
\newblock \emph{European Journal of Social Psychology}, 8\penalty0
  (3):\penalty0 401--405, 1978.

\bibitem[Friedkin et~al.(2019)Friedkin, Proskurnikov, and
  Bullo]{friedkin2019positive}
Noah~E Friedkin, Anton~V Proskurnikov, and Francesco Bullo.
\newblock Positive contagion and the macrostructures of generalized balance.
\newblock \emph{Network Science}, pages 1--14, 2019.

\bibitem[Harary(1959)]{harary1959measurement}
Frank Harary.
\newblock On the measurement of structural balance.
\newblock \emph{Behavioral Science}, 4\penalty0 (4):\penalty0 316--323, 1959.

\bibitem[Abell(1968)]{abell1968structural}
Peter Abell.
\newblock Structural balance in dynamic structures.
\newblock \emph{Sociology}, 2\penalty0 (3):\penalty0 333--352, 1968.

\bibitem[de~Nooy(1999)]{de1999sign}
Wouter de~Nooy.
\newblock The sign of affection: Balance-theoretic models and incomplete signed
  digraphs.
\newblock \emph{Social Networks}, 21\penalty0 (3):\penalty0 269--286, 1999.

\bibitem[Kunegis et~al.(2010)Kunegis, Schmidt, Lommatzsch, Lerner, De~Luca, and
  Albayrak]{kunegis2010spectral}
J{\'e}r{\^o}me Kunegis, Stephan Schmidt, Andreas Lommatzsch, J{\"u}rgen Lerner,
  Ernesto~W De~Luca, and Sahin Albayrak.
\newblock Spectral analysis of signed graphs for clustering, prediction and
  visualization.
\newblock In \emph{Proceedings of the 2010 SIAM International Conference on
  Data Mining}, pages 559--570, 2010.

\bibitem[Terzi and Winkler(2011)]{terzi2011spectral}
Evimaria Terzi and Marco Winkler.
\newblock A spectral algorithm for computing social balance.
\newblock In \emph{International Workshop on Algorithms and Models for the
  Web-Graph}, pages 1--13. Springer, 2011.

\bibitem[Facchetti et~al.(2011)Facchetti, Iacono, and
  Altafini]{facchetti2011computing}
Giuseppe Facchetti, Giovanni Iacono, and Claudio Altafini.
\newblock Computing global structural balance in large-scale signed social
  networks.
\newblock \emph{Proceedings of the National Academy of Sciences}, 108\penalty0
  (52):\penalty0 20953--20958, 2011.

\bibitem[Rawlings and Friedkin(2017)]{rawlings2017structural}
Craig~M Rawlings and Noah~E Friedkin.
\newblock The structural balance theory of sentiment networks: Elaboration and
  test.
\newblock \emph{American Journal of Sociology}, 123\penalty0 (2):\penalty0
  510--548, 2017.
\newblock \doi{10.1086/692757}.
\newblock URL \url{https://doi.org/10.1086/692757}.

\bibitem[Szell et~al.(2010)Szell, Lambiotte, and Thurner]{szell2010}
Michael Szell, Renaud Lambiotte, and Stefan Thurner.
\newblock Multirelational organization of large-scale social networks in an
  online world.
\newblock \emph{Proceedings of the National Academy of Sciences}, 107\penalty0
  (31):\penalty0 13636--13641, 2010.
\newblock ISSN 0027-8424.
\newblock \doi{10.1073/pnas.1004008107}.
\newblock URL \url{https://www.pnas.org/content/107/31/13636}.

\bibitem[Askarisichani et~al.(2019)Askarisichani, Lane, Bullo, Friedkin, Singh,
  and Uzzi]{nc2019}
Omid Askarisichani, Jacqueline~Ng Lane, Francesco Bullo, Noah~E Friedkin,
  Ambuj~K Singh, and Brian Uzzi.
\newblock Structural balance emerges and explains performance in risky
  decision-making.
\newblock \emph{Nature communications}, 10, 2019.

\bibitem[Leskovec et~al.(2010)Leskovec, Huttenlocher, and
  Kleinberg]{leskovec2010predicting}
Jure Leskovec, Daniel Huttenlocher, and Jon Kleinberg.
\newblock Predicting positive and negative links in online social networks.
\newblock In \emph{Proceedings of the 19th International Conference on World
  Wide Web}, pages 641--650. ACM, 2010.

\bibitem[Newcomb(1956)]{newcomb1956prediction}
Theodore~M Newcomb.
\newblock The prediction of interpersonal attraction.
\newblock \emph{American Psychologist}, 11\penalty0 (11):\penalty0 575, 1956.

\bibitem[Shahriari et~al.(2016)Shahriari, Sichani, Gharibshah, and
  Jalili]{shahriari2016sign}
Mohsen Shahriari, Omid~Askarisichani Sichani, Joobin Gharibshah, and Mahdi
  Jalili.
\newblock Sign prediction in social networks based on users reputation and
  optimism.
\newblock \emph{Social Network Analysis and Mining}, 6\penalty0 (1):\penalty0
  91, 2016.

\bibitem[Lazer et~al.(2009)Lazer, Pentland, Adamic, Aral, Barab{\'a}si, Brewer,
  Christakis, Contractor, Fowler, Gutmann, et~al.]{lazer2009computational}
David Lazer, Alex Pentland, Lada Adamic, Sinan Aral, Albert-L{\'a}szl{\'o}
  Barab{\'a}si, Devon Brewer, Nicholas Christakis, Noshir Contractor, James
  Fowler, Myron Gutmann, et~al.
\newblock Computational social science.
\newblock \emph{Science}, 323\penalty0 (5915):\penalty0 721--723, 2009.

\bibitem[Shilliday and Lautenschlager(2012)]{shilliday2012data}
Andrew Shilliday and Jennifer Lautenschlager.
\newblock Data for a worldwide {ICEWS} and ongoing research.
\newblock \emph{Advances in Design for Cross-Cultural Activities}, page 455,
  2012.

\bibitem[Shils et~al.(1975)]{shils1975center}
Edward Shils et~al.
\newblock \emph{Center and periphery}.
\newblock Chicago: University of Chicago Press, 1975.

\bibitem[Bourgeois and Friedkin(2001)]{bourgeois2001distant}
Michael Bourgeois and Noah~E Friedkin.
\newblock The distant core: social solidarity, social distance and
  interpersonal ties in core--periphery structures.
\newblock \emph{Social networks}, 23\penalty0 (4):\penalty0 245--260, 2001.

\bibitem[S{\o}rensen and Hallinan(1976)]{sorensen1976stochastic}
Aage~B S{\o}rensen and Maureen~T Hallinan.
\newblock A stochastic model for change in group structure.
\newblock \emph{Social Science Research}, 5\penalty0 (1):\penalty0 43--61,
  1976.

\bibitem[Marvel et~al.(2011)Marvel, Kleinberg, Kleinberg, and
  Strogatz]{marvel2011}
Seth~A. Marvel, Jon Kleinberg, Robert~D. Kleinberg, and Steven~H. Strogatz.
\newblock Continuous-time model of structural balance.
\newblock \emph{Proceedings of the National Academy of Sciences}, 108\penalty0
  (5):\penalty0 1771--1776, 2011.
\newblock ISSN 0027-8424.
\newblock \doi{10.1073/pnas.1013213108}.
\newblock URL \url{https://www.pnas.org/content/108/5/1771}.

\bibitem[Davis(1967)]{davis1967clustering}
James~A Davis.
\newblock Clustering and structural balance in graphs.
\newblock \emph{Human Relations}, 20\penalty0 (2):\penalty0 181--187, 1967.

\bibitem[Holland and Leinhardt(1971)]{holland1971transitivity}
Paul~W Holland and Samuel Leinhardt.
\newblock Transitivity in structural models of small groups.
\newblock \emph{Comparative Group Studies}, 2\penalty0 (2):\penalty0 107--124,
  1971.

\bibitem[Kumar et~al.(2016)Kumar, Spezzano, Subrahmanian, and
  Faloutsos]{kumar2016edge}
Srijan Kumar, Francesca Spezzano, VS~Subrahmanian, and Christos Faloutsos.
\newblock Edge weight prediction in weighted signed networks.
\newblock In \emph{IEEE International Conference on Data Mining}, pages
  221--230, 2016.

\bibitem[Kumar et~al.(2018)Kumar, Hooi, Makhija, Kumar, Faloutsos, and
  Subrahmanian]{kumar2018rev2}
Srijan Kumar, Bryan Hooi, Disha Makhija, Mohit Kumar, Christos Faloutsos, and
  VS~Subrahmanian.
\newblock Rev2: Fraudulent user prediction in rating platforms.
\newblock In \emph{Proceedings of the ACM International Conference on Web
  Search and Data Mining}, pages 333--341. ACM, 2018.

\bibitem[Barbieri et~al.(2009)Barbieri, Keshk, and
  Pollins]{barbieri2009trading}
Katherine Barbieri, Omar~MG Keshk, and Brian~M Pollins.
\newblock Trading data: Evaluating our assumptions and coding rules.
\newblock \emph{Conflict Management and Peace Science}, 26\penalty0
  (5):\penalty0 471--491, 2009.

\bibitem[Martin et~al.(2008)Martin, Mayer, and Thoenig]{martin2008make}
Philippe Martin, Thierry Mayer, and Mathias Thoenig.
\newblock Make trade not war?
\newblock \emph{The Review of Economic Studies}, 75\penalty0 (3):\penalty0
  865--900, 2008.

\bibitem[Oneal and Russett(1999)]{oneal1999assessing}
John~R Oneal and Bruce Russett.
\newblock Assessing the liberal peace with alternative specifications: Trade
  still reduces conflict.
\newblock \emph{Journal of Peace Research}, 36\penalty0 (4):\penalty0 423--442,
  1999.

\bibitem[Hegre et~al.(2010)Hegre, Oneal, and Russett]{hegre2010trade}
H{\aa}vard Hegre, John~R Oneal, and Bruce Russett.
\newblock Trade does promote peace: New simultaneous estimates of the
  reciprocal effects of trade and conflict.
\newblock \emph{Journal of Peace Research}, 47\penalty0 (6):\penalty0 763--774,
  2010.

\bibitem[Granger(1988)]{granger1988causality}
Clive~WJ Granger.
\newblock Causality, cointegration, and control.
\newblock \emph{Journal of Economic Dynamics and Control}, 12\penalty0
  (2-3):\penalty0 551--559, 1988.

\bibitem[Friedkin(2011)]{friedkin2011formal}
Noah~E Friedkin.
\newblock A formal theory of reflected appraisals in the evolution of power.
\newblock \emph{Administrative Science Quarterly}, 56\penalty0 (4):\penalty0
  501--529, 2011.

\bibitem[Heider(2013)]{heider2013psychology}
Fritz Heider.
\newblock \emph{The Psychology of Interpersonal Relations}.
\newblock Psychology Press, 2013.

\bibitem[Johnsen(1985)]{johnsen1985network}
Eugene~C Johnsen.
\newblock Network macrostructure models for the {Davis-Leinhardt} set of
  empirical sociomatrices.
\newblock \emph{Social Networks}, 7\penalty0 (3):\penalty0 203--224, 1985.

\bibitem[Montgomery(2009)]{montgomery2009balance}
James~D Montgomery.
\newblock Balance theory with incomplete awareness.
\newblock \emph{Journal of Mathematical Sociology}, 33\penalty0 (2):\penalty0
  69--96, 2009.

\bibitem[Chiba et~al.(2017)Chiba, Hino, Akaho, and Murata]{chiba2017time}
Tomoaki Chiba, Hideitsu Hino, Shotaro Akaho, and Noboru Murata.
\newblock Time-varying transition probability matrix estimation and its
  application to brand share analysis.
\newblock \emph{PLOS One}, 12\penalty0 (1):\penalty0 e0169981, 2017.

\bibitem[Friedman et~al.(2010)Friedman, Hastie, and
  Tibshirani]{friedman2010note}
Jerome Friedman, Trevor Hastie, and Robert Tibshirani.
\newblock A note on the group lasso and a sparse group lasso.
\newblock \emph{arXiv preprint arXiv:1001.0736}, 2010.

\bibitem[Boyd and Vandenberghe(2004)]{boyd2004convex}
Stephen Boyd and Lieven Vandenberghe.
\newblock \emph{Convex Optimization}.
\newblock Cambridge University Press, 2004.

\bibitem[Diamond and Boyd(2016)]{cvxpy}
Steven Diamond and Stephen Boyd.
\newblock {CVXPY}: A {P}ython-embedded modeling language for convex
  optimization.
\newblock \emph{Journal of Machine Learning Research}, 17\penalty0
  (83):\penalty0 1--5, 2016.

\bibitem[Boschee et~al.(2015)Boschee, Lautenschlager, O’Brien, Shellman,
  Starz, and Ward]{boschee2015icews}
Elizabeth Boschee, Jennifer Lautenschlager, Sean O’Brien, Steve Shellman,
  James Starz, and Michael Ward.
\newblock {ICEWS} coded event data.
\newblock \emph{Harvard Dataverse}, 12, 2015.

\bibitem[Mammen et~al.(1997)Mammen, van~de Geer, et~al.]{mammen1997locally}
Enno Mammen, Sara van~de Geer, et~al.
\newblock Locally adaptive regression splines.
\newblock \emph{The Annals of Statistics}, 25\penalty0 (1):\penalty0 387--413,
  1997.

\bibitem[Tibshirani et~al.(2014)]{tibshirani2014adaptive}
Ryan~J Tibshirani et~al.
\newblock Adaptive piecewise polynomial estimation via trend filtering.
\newblock \emph{The Annals of Statistics}, 42\penalty0 (1):\penalty0 285--323,
  2014.

\bibitem[Kim et~al.(2009)Kim, Koh, Boyd, and Gorinevsky]{kim2009ell_1}
Seung-Jean Kim, Kwangmoo Koh, Stephen Boyd, and Dimitry Gorinevsky.
\newblock $\backslash$ell\_1 trend filtering.
\newblock \emph{SIAM review}, 51\penalty0 (2):\penalty0 339--360, 2009.

\bibitem[Wang et~al.(2016)Wang, Sharpnack, Smola, and
  Tibshirani]{wang2016trend}
Yu-Xiang Wang, James Sharpnack, Alexander~J Smola, and Ryan~J Tibshirani.
\newblock Trend filtering on graphs.
\newblock \emph{The Journal of Machine Learning Research}, 17\penalty0
  (1):\penalty0 3651--3691, 2016.

\end{thebibliography}

\newpage
\onecolumn
\begin{huge}
\textbf{Appendix} 
\end{huge}
\section{Results on other datasets}

To buttress the finding of high transition probability toward and staying in more balanced triads, we apply our experiments on two other datasets. We report the results of the time-varying Markov model on estimating the underlying probability transitions for all three datasets: ICEWS, Bitcoin Alpha, and Bitcoin OTC.

\paragraph{Bitcoin Alpha trust weighted signed network}
This is a who-trusts-whom network of people who trade using Bitcoin on a platform called Bitcoin Alpha. Since Bitcoin users are anonymous, there is a need to maintain a record of users' reputations to prevent transactions with fraudulent and risky users. Members of Bitcoin Alpha rate other members on a scale of -10 (total distrust) to +10 (total trust)~\cite{kumar2016edge, kumar2018rev2}.

\paragraph{Bitcoin OTC trust weighted signed network:}
Very similar to the previous dataset, Bitcoin OTC is another platform for trading Bitcoin and has similar trust edges over time~\cite{kumar2016edge, kumar2018rev2}.

The statistics about all datasets are depicted in Table~\ref{tbl:datasets_stats}.

\begin{table}[H]
    \centering
    \caption{Brief statistics about the datasets used in this study.}
    \begin{tabular}{ll}
    
    \midrule
    Name & \textbf{Integrated Crisis Early Warning System (ICEWS)} \\
    \#Nodes & 250 \\
    \#Edges & 8,073,921 \\
    \#Positive edges & 32,029 (90\%) \\
    \#Negative edges & 3,563 (10\%) \\
    Edge weight & -10 to +10\\
    Spans through & 1995-01-01 to 2018-09-30 (23+ years)\\\\
    
    \midrule
    Name & \textbf{Bitcoin Alpha} \\
    \#Nodes & 3,783 \\
    \#Edges & 24,186 \\
    \#Positive edges & 22,650 (94\%) \\
    \#Negative edges & 1,536 (6\%) \\
    Edge weight & -10 to +10\\
    Spans through & 2010-11-07 to 2016-01-21 (5+ years)\\\\
    
    \midrule
    Name & \textbf{Bitcoin OTC} \\
    \#Nodes & 5,881 \\
    \#Edges & 35,592 \\
    \#Positive edges & 32,029 (90\%) \\
    \#Negative edges & 3,563 (10\%) \\
    Edge weight & -10 to +10\\
    Spans through & 2010-11-08 to 2016-01-24 (5+ years)\\\\
    \bottomrule
    
    \end{tabular}
    \label{tbl:datasets_stats}
\end{table}

\begin{figure}[H]
    \centering
    \xincludegraphics[width=0.47\textwidth, label=\textbf{a)}]
    {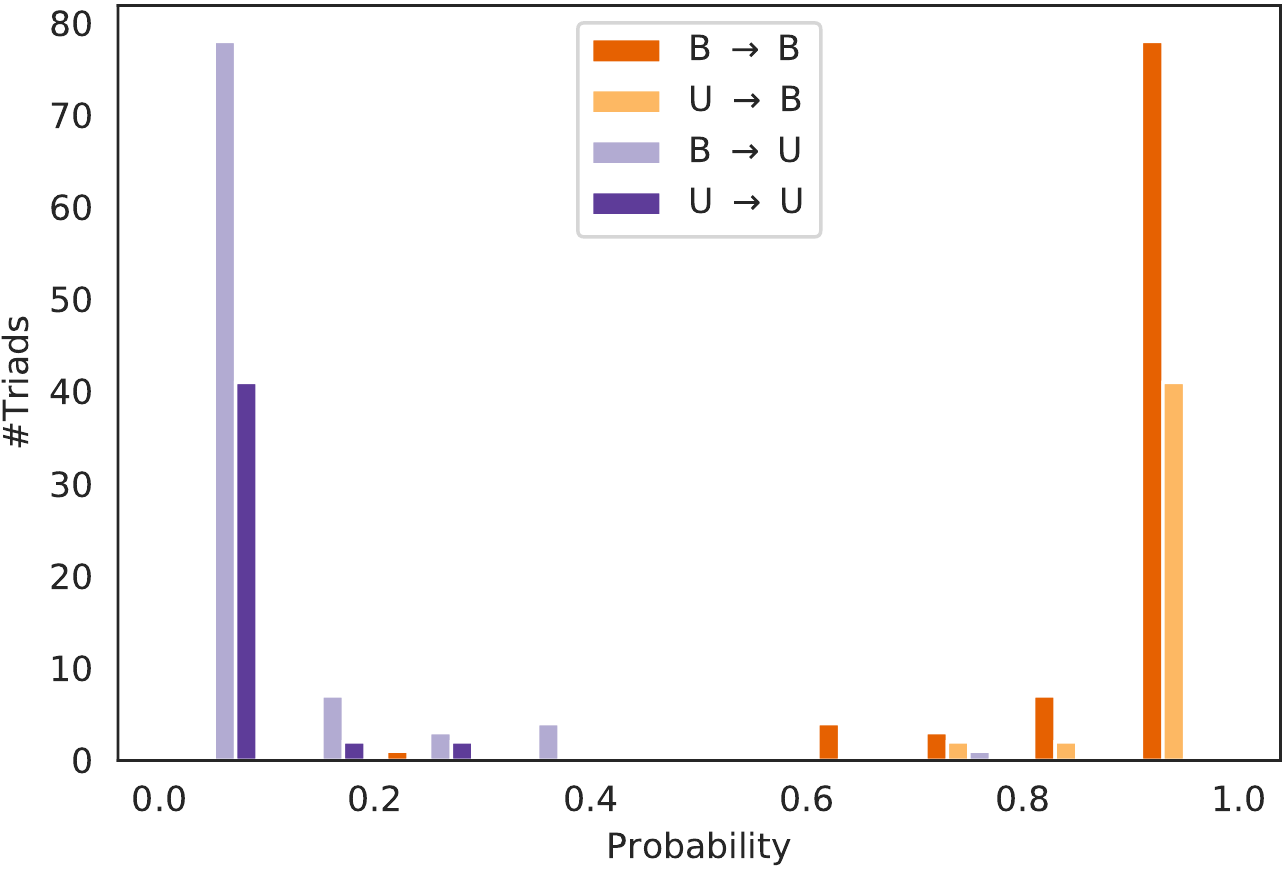}
    \xincludegraphics[width=0.47\textwidth, label=\textbf{b)}]
    {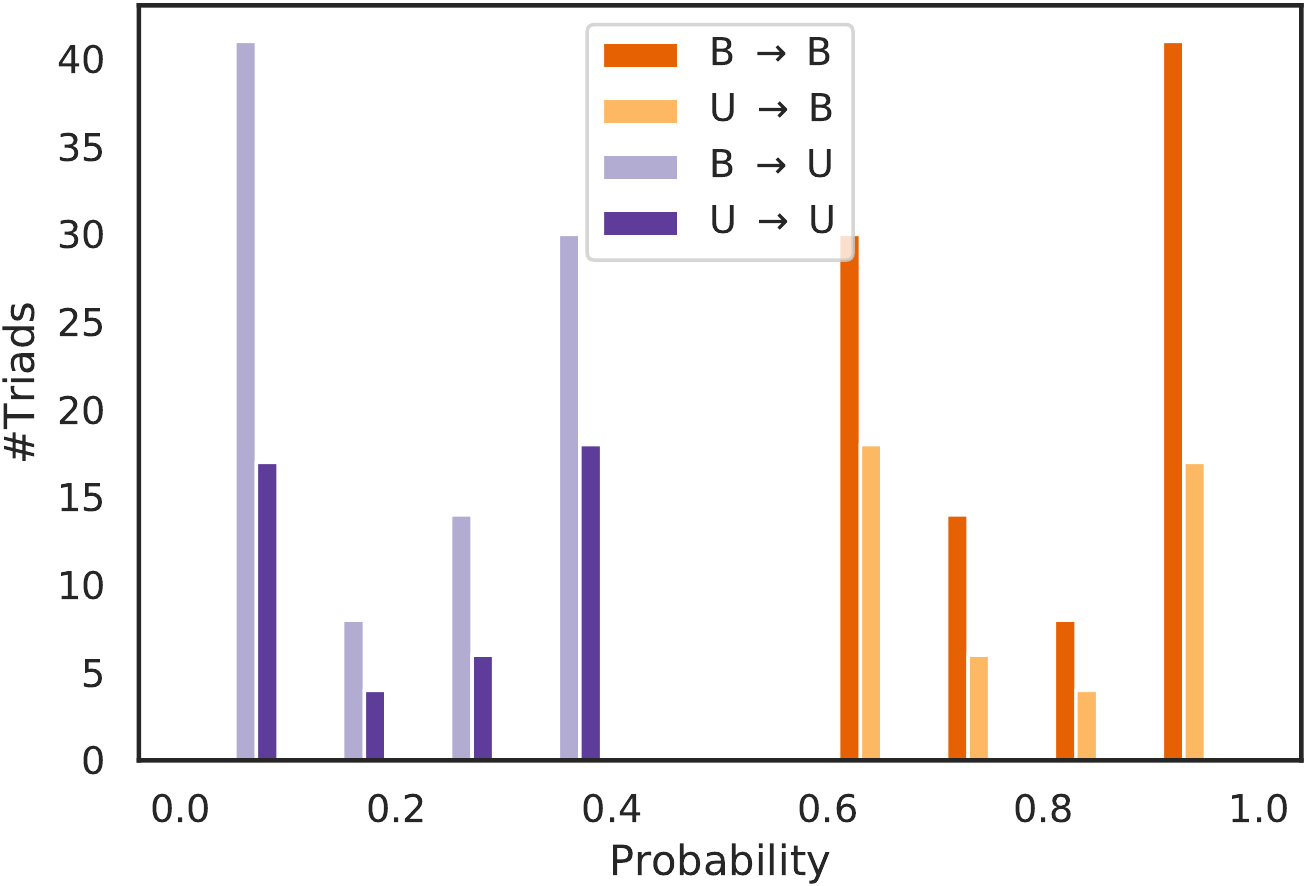}
    \xincludegraphics[width=0.47\textwidth, label=\textbf{c)}]
    {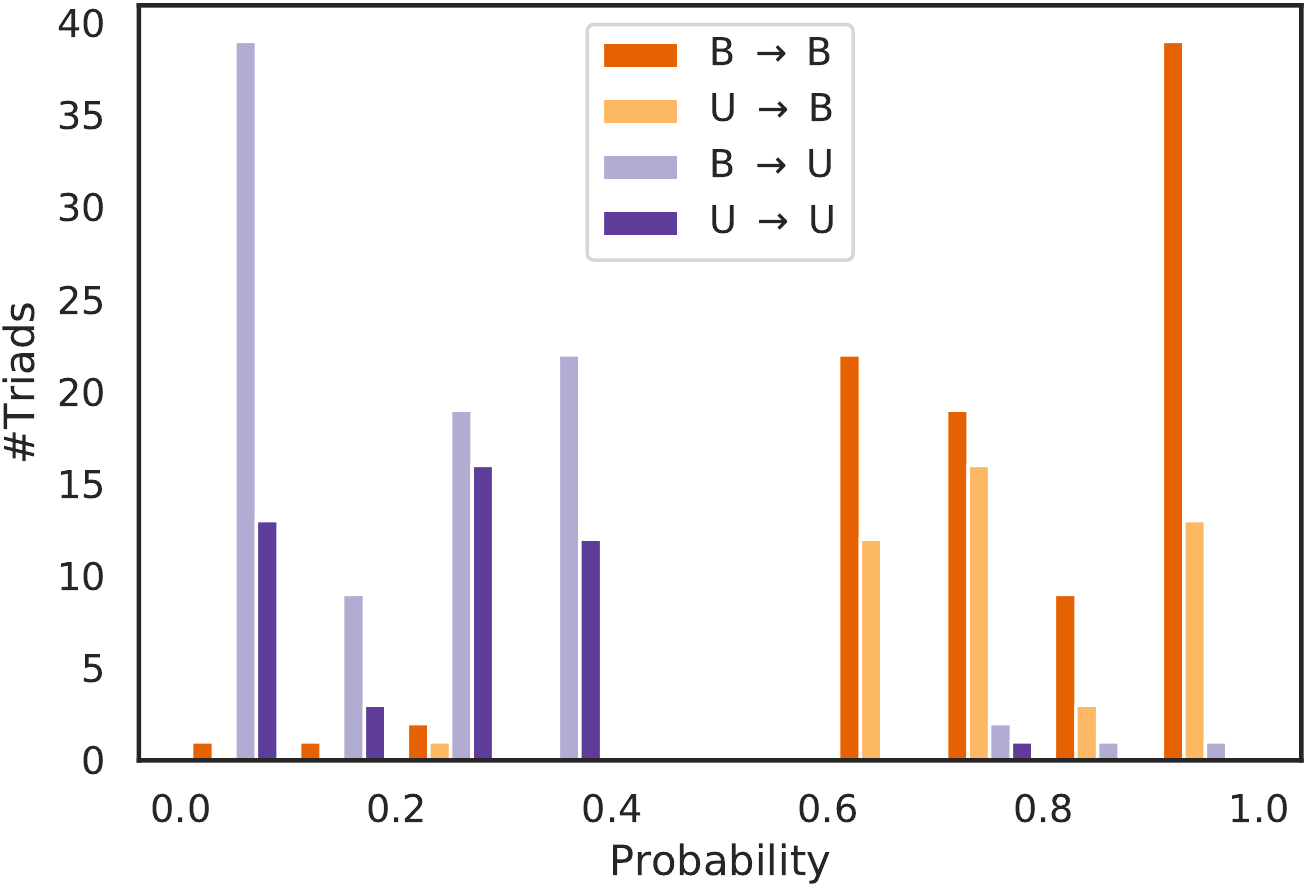}
    \caption{The distribution of transition probability within the quadrants of transitivity-balance, in the estimated Markov transition matrix, for dataset a) ICEWS, b) Bitcoin Alpha, c) Bitcoin OTC. In the legend, B means balance, U means unbalanced, and for instance U $\rightarrow$ B shows the probability transition of unbalanced triad types to balanced triad types. The figure for clustering- and classical- of structural balance depict similar findings. The Markov transition matrix is estimated by solving ~\eqref{eq:objective_function} and the analysis pipeline is described in Fig. 4 in the main paper.}
    \label{fig:balance_transition_probabilities}
\end{figure}

Fig.~\ref{fig:boxplot_transition_probabilities} shows the evidence for the network dynamic toward balance for these three datasets. The probability of transitions from unbalanced to balanced triads is significantly higher than transitions from balanced to unbalanced triads. The probability of remaining balanced is more likely than the probability remaining unbalanced. These findings are most strongly expressed in the data on transitivity-balance. This especially strong expression of transitivity-driven evolution is consistent with its status as the most important axiom of structural balance theory~\cite{heider1946attitudes, holland1971transitivity}.
 
Interestingly, Fig.~\ref{fig:boxplot_transition_probabilities} shows that transition toward and staying in structurally balance hold regardless of the definition of balance (in the main paper) and the setting (international or financial networks), but notably, they have the strongest expression with transitivity-balance.

\begin{figure}[H]
\begin{center}
\begin{tabular}{llllr}
& \multicolumn{1}{c}{ICEWS} & \multicolumn{1}{c}{Bitcoin Alpha} & \multicolumn{1}{c}{Bitcoin OTC} & \\

\begin{tabular}[c]{@{}l@{}}balanced $\rightarrow$unbalanced\\ unbalanced $\rightarrow$unbalanced\\ unbalanced $\rightarrow$balanced\\ balanced $\rightarrow$balanced\end{tabular} & \adjustimage{width=0.2\textwidth, valign=m}{ICEWS_transitivity_nolabels}         & \adjustimage{width=0.2\textwidth, valign=m}{Bitcoin_Alpha_transitivity_nolabels}         & \adjustimage{width=0.2\textwidth, valign=m}{Bitcoin_OTC_transitivity_nolabels} & transitivity \\\\\\

\begin{tabular}[c]{@{}l@{}}balanced $\rightarrow$unbalanced\\ unbalanced $\rightarrow$unbalanced\\ unbalanced $\rightarrow$balanced\\ balanced $\rightarrow$balanced\end{tabular} & \adjustimage{width=0.2\textwidth, valign=m}{ICEWS_clustering_nolabels} & \adjustimage{width=0.2\textwidth, valign=m}{Bitcoin_Alpha_clustering_nolabels} & \adjustimage{width=0.2\textwidth, valign=m}{Bitcoin_OTC_clustering_nolabels} & clustering \\\\\\

\begin{tabular}[c]{@{}l@{}}balanced $\rightarrow$unbalanced\\ unbalanced $\rightarrow$unbalanced\\ unbalanced $\rightarrow$balanced\\ balanced $\rightarrow$balanced\end{tabular} & \adjustimage{width=0.2\textwidth, valign=m}{ICEWS_classical_nolabels}   & \adjustimage{width=0.2\textwidth, valign=m}{Bitcoin_Alpha_classical_nolabels}   & \adjustimage{width=0.2\textwidth, valign=m}{Bitcoin_OTC_classical_nolabels} & classical

\end{tabular}
\end{center}
\caption{Estimated transition probability for classical-, clustering-, and transitivity-balanced and -unbalanced triads in all three datasets (the pipeline is described in Fig. 4 in the main paper and datasets in Table~\ref{tbl:datasets_stats}). The x-axis shows the estimated probability (computed by solving ~\eqref{eq:objective_function}), the box is the interquartile range of the probability distribution, the orange line is the median of the distribution, and the whisker shows minimum and maximum of the range of the distribution. This figure shows that the probability of transitions from unbalanced triads to balanced ones is significantly higher than the opposite transitions. Also, the probability of remaining balanced is more likely than the probability of remaining unbalanced. Surprisingly, these findings are robust with respect of multiple definitions of balance and different settings.}
\label{fig:boxplot_transition_probabilities_full}
\end{figure}

\begin{figure}[H]
    \centering
    \includegraphics[width=1.0\linewidth]{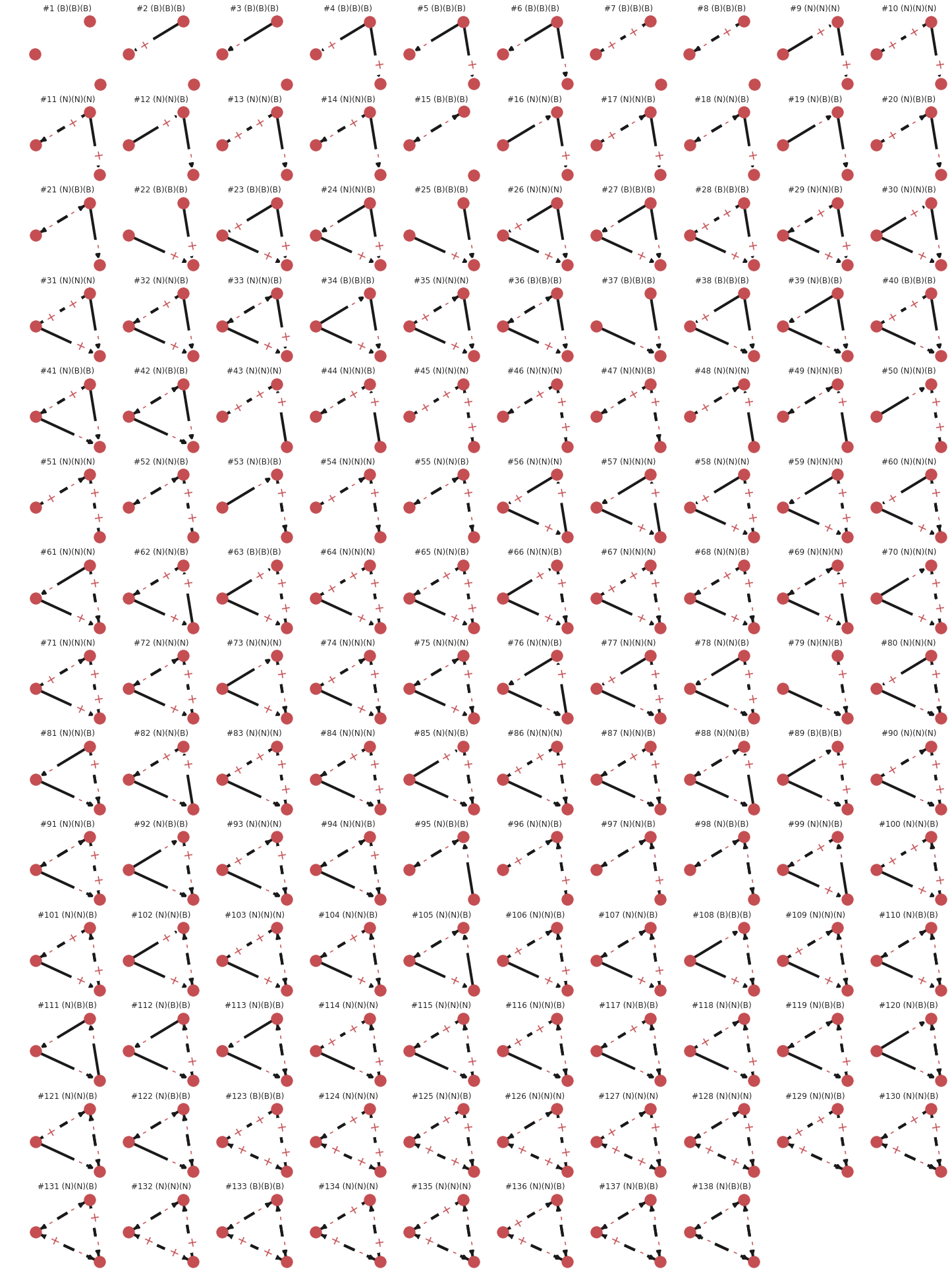}
    \caption{All possible 138 sparse triads. Title of each triad shows its \#ID, and whether if it is balanced with respect to classical-, clustering-, and transitivity-balanced, respectively. The title includes (B) if it is balanced, and (N) if not, for every definition. Notably, when nodes are aware of every other one, 24 triads out of 138 ones are classically-balanced. 44 triads out of 138 are clustering-balanced. Also, 93 triads out of 138 are transitivity-balanced (see the definitions in the main paper).}
    \label{fig:triads}
\end{figure}

\section{Proofs for the time-varying Markov model}
\paragraph{Settings:}

There are $n=138$ states, named $l_1, \dots, l_{n}$. There are $m$ entities (triads in the network), that in parallel, change states for $T + 1$ periods of time. Each empirical Markov probability transition matrix, $\hat P_t$, is computed as follows

\begin{equation}
\hat P_t = \frac{\sum\limits_{i=1}^{m} \indicator(S_t^{(i)}=l_t,\; S_{t-1}^{(i)}=l_{t-1})}{\sum\limits_{i=1}^{m} \indicator(S_{t-1}^{(i)}=l_{t-1})}
\end{equation}

where each $\hat P_t$ is a $n \times n$ matrix and there exist $\hat P_1, \dots, \hat P_T$. $S_t$ shows the state of triad $i$ (three countries) at time $t$.

\paragraph{Assumptions:}
The assumption is there is an error in empirical transition matrices such that

\begin{equation}\label{eq:assumption}
    \hat P_t = P_{0, t} + z_t
\end{equation}

where $P_{0, t}: t = 1, \dots, T$ are the true unknown transition matrices, $\hat P_{t}, t = 1, \dots, T$ the empirical transition matrices, and $z_t, t = 1, \dots, T$ are i.i.d sub-Gaussian errors with zero mean. The probability transitions are between $[0, 1]$; thus, it is easy to show that the error is bounded as there exists a value for $M, \sigma > 0$ such that

$$P(|z_i| > t)\leqwithspaces M \exp(-t^2 / (2\sigma^2))\;\; \forall t > 0,\; i=1,\dots,n.$$

Hence, $z_t$ is sub-Gaussian. In our results, we also find empirically support for independence of errors as the Pearson correlation of every cells for subsequent estimated matrices are very small and more than 70\% are not statistically significant ($p \geq 0.05$).

We also assume the total variation~\cite{mammen1997locally} of matrices does not grow too quickly \cite{tibshirani2014adaptive}, where each $P_{0, t}$ is a matrix with $n^2$ dimensions, for the constant value of $n = 138$

\begin{equation}\label{eq:growing_total_variation_condition}
    TV(P_{0, t}) = \sum_{i=2}^T ||P_{0, t} - P_{0, t-1}||_{1,1} \leq n^2 C_T = \bigo(T).
\end{equation}

We empirically report the above equation for our data, is indeed $\bigo(T)$ and precisely is $0.04T$.

\paragraph{Problem definition:}
Instead of having a model for one Markov transition matrix and fit that to the entire period, we define a convex optimization problem to predict all transition matrices altogether using trend filtering for nonparametric regression \cite{kim2009ell_1}.

\begin{equation}\label{eq:obj_function}
    \begin{array}{lll}
        \widetilde{P_1}, \dots, \widetilde{P_T} & = \minimize\limits_{P_1, \dots, P_T} & \halft \sumtot ||\hat P_t - P_t||_F^2 + \lambda_1 \sumtwotot ||P_t - P_{t-1}||_{1, 1} + \lambda_2 \sumtwotot ||P_t - P_{t-1}||_{2, 1}, \\
        & \subjectto & (P_t)_{ij} > 0,\; \forall i, j \in [1, n]\\
        &           & \ones^T P_t = \ones^T
    \end{array}
\end{equation}

where $\widetilde{P_t}: t = 1, \dots, T$ are the estimated transition matrices. $\lambda_1$ and $\lambda_2$ are the hyperparameters which are tuned by applying Grid Search with 5-fold cross-validation.

\subsection{Convexity proof}
The objective function is a summation of three norms which are convex, all of the inequality constraints are convex, and all equality constraints are affine~\cite{boyd2004convex}. Therefore, the problem is convex, it has a globally optimal solution~\cite{boyd2004convex}. Thus, we solve this equation by a convex optimization solver, CVXPY~\cite{cvxpy}.

\subsection{Convergence Rate Proof}
\paragraph{Convergence Rate Theorem:}
\begin{equation}
	\sumtot ||\widetilde{P_t} - P_{0, t}||_F^2 = \bigoprob\bigg(n^2 (\text{nullity}(\Delta) + M \sqrt{\log r} C_T)\bigg).
\end{equation}

\begin{proof}

Since the objective function in ~\eqref{eq:obj_function}, in previous proof is shown to be convex; thanks to the optimiality of {\em argmin}, $\widetilde{P}$, the solution of the optimization problem minimizes the objective function more than any other matrix, say $X$,

$$\text{objective}(\widetilde{P})\leqwithspaces\text{objective}(X).$$

We use ~\eqref{eq:obj_function} and rewrite it as

$$\halft\sumtot ||\hat P_t - \widetilde{P_t}||_F^2 + \lambda_1 \sumtwotot ||\widetilde{P_t} - \widetilde{P}_{t-1}||_{1, 1}\leqwithspaces\halft\sumtot ||\hat P_t - X_t||_F^2 + \lambda_1 \sumtwotot ||X_t - X_{t-1}||_{1, 1}.$$

As a matter of fact $X_t$, could be replaced by $P_{0, t}$ as follows

$$\halft\sumtot ||\hat P_t - \widetilde{P_t}||_F^2 + \lambda_1 \sumtwotot ||\widetilde{P_t} - \widetilde{P}_{t-1}||_{1, 1}\leqwithspaces\halft\sumtot ||\hat P_t - P_{0, t}||_F^2 + \lambda_1 \sumtwotot ||P_{0, t} - P_{0, t-1}||_{1, 1}.$$

After multiplying both sides by 2 and expanding the previous inequality by using the assumption in ~\eqref{eq:assumption}, we have

$$\sumtot ||P_{0, t} + z_t - \widetilde{P_t}||_F^2 + 2\lambda_1 T \sumtwotot ||\widetilde{P_t} - \widetilde{P}_{t-1}||_{1, 1}\leqwithspaces\sumtot || P_{0, t} + z_t - P_{0, t}||_F^2 + 2\lambda_1 T \sumtwotot ||P_{0, t} - P_{0, t-1}||_{1, 1}.$$

Then, we can write

$$\sumtot ||(P_{0, t} - \widetilde{P_t}) + z_t)||_F^2 + 2\lambda_1 T \sumtwotot ||\widetilde{P_t} - \widetilde{P}_{t-1}||_{1, 1}\leqwithspaces\sumtot ||z_t||_F^2 + 2\lambda_1 T \sumtwotot ||P_{0, t} - P_{0, t-1}||_{1, 1}.$$

By expanding the power of two in the left most term in the above inequality, we have

\begin{align*}
    \sumtot ||P_{0, t} - \widetilde{P_t}||_F^2 + \sumtot ||z_t||_F^2 + 2\sumtot z_t^T(P_{0, t} - \widetilde{P_t}) + 2\lambda_1 T \sumtwotot ||\widetilde{P_t} - \widetilde{P}_{t-1}||_{1, 1}\\ \leqwithspaces\sumtot ||z_t||_F^2 + 2\lambda_1 T \sumtwotot ||P_{0, t} - P_{0, t-1}||_{1, 1},
\end{align*}

where rearranging the terms yields

$$\sumtot ||\widetilde{P_t} - P_{0, t}||_F^2\leqwithspaces 2\sumtot z_t^T(\widetilde{P_t} - P_{0, t}) + 2\lambda_1 T \sumtwotot ||P_{0, t} - P_{0, t-1}||_{1, 1} - 2\lambda_1 T \sumtwotot ||\widetilde{P_t} - \widetilde{P}_{t-1}||_{1, 1}.$$

Using orthogonal decomposition on the left term we have

\begin{equation}\label{eq:decomposition}
    \sumtot ||\widetilde{P_t} - P_{0, t}||_F^2 = \sumtot ||z_t||_{R_{\bot}}^2 + \sumtot ||\widetilde{P_t} - P_{0, t}||_R^2,
\end{equation}

where the null space term is of the order

\begin{equation}\label{eq:null_space}
    \sumtot ||z_t||_{R_{\bot}}^2 = \bigo(n^2 \text{nullity}(\Delta)),
\end{equation}

and the row space term of inequality ~~\eqref{eq:decomposition} is rewritten as following

$$\sumtot ||\widetilde{P_t} - P_{0, t}||_R^2\leqwithspaces 2\sumtot z_t^T P_R (\widetilde{P_t} - P_{0, t}) + 2\lambda_1 T \sumtwotot ||P_{0, t} - P_{0, t-1}||_{1, 1} - 2\lambda_1 T \sumtwotot ||\widetilde{P_t} - \widetilde{P}_{t-1}||_{1, 1}.$$

In the first term we use $P_R = \Delta^\dagger\Delta$ where $\Delta \in \mathbb{R}^{r \times n}$ is an arbitrary linear operator, with $r$ rows. We have

\begin{equation}\label{eq:before_holder}
    \sumtot ||\widetilde{P_t} - P_{0, t}||_R^2\leqwithspaces 2\sumtot z_t^T \Delta^\dagger\Delta (\widetilde{P_t} - P_{0, t}) + 2\lambda_1 T \sumtwotot ||P_{0, t} - P_{0, t-1}||_{1, 1} - 2\lambda_1 T \sumtwotot ||\widetilde{P_t} - \widetilde{P}_{t-1}||_{1, 1}.
\end{equation}

Based on H\"{o}lder's inequality, we know for any $p, q \geq 1$ such that $\frac 1 p + \frac 1 q = 1$ then for any two functions $f$ and $g$ the following inequality always holds

$$\|fg\|_{1}\leqwithspaces\|f\|_{p}\|g\|_{q}.$$

By applying H\"{o}lder's inequality on the term $z_t^T \Delta^\dagger\Delta (\widetilde{P_t} - P_{0, t})$ from ~\eqref{eq:before_holder}, with $p=\infty$ and $q=1$, we see

$$z_t^T \Delta^\dagger\Delta (\widetilde{P_t} - P_{0, t})\leqwithspaces ||z_t^T \Delta^\dagger\Delta (\widetilde{P_t} - P_{0, t})||_1\leqwithspaces ||(\Delta^\dagger)^T z_t||_\infty ||\Delta (\widetilde{P_t} - P_{0, t})||_1,$$

in which $\Delta (\widetilde{P_t} - P_{0, t}) = (\widetilde{P}_t - \widetilde{P}_{t-1}) - (P_{0, t} - P_{0, t-1})$ and we can rewrite the inequality as

\begin{equation}\label{eq:holder_mid}
    z_t^T \Delta^\dagger\Delta (\widetilde{P_t} - P_{0, t})\leqwithspaces ||(\Delta^\dagger)^T z_t||_\infty ||(\widetilde{P}_t - \widetilde{P}_{t-1}) - (P_{0, t} - P_{0, t-1})||_1.
\end{equation}

We claim that variation in every step from its previous value has the same sign in the ground-truth and the estimated matrix. In other words, sign of $P_{0, t} - P_{0, t-1}$ and $\widetilde{P}_t - \widetilde{P}_{t-1}$ is always the same. Thus, we can use the inequality $||x - y|| \leq ||x + y||$ where $xy \geq 0$. Therefore, we have

$$z_t^T \Delta^\dagger\Delta (\widetilde{P_t} - P_{0, t})\leqwithspaces ||(\widetilde{P}_t - \widetilde{P}_{t-1}) - (P_{0, t} - P_{0, t-1})||_1 \leqwithspaces ||(\widetilde{P}_t - \widetilde{P}_{t-1}) + (P_{0, t} - P_{0, t-1})||_1$$

Based on triangle inequality ($||x + y||_1 \leq ||x||_1 + ||y||_1$ for any $x, y$), we know the right term in previous equation is

\begin{align}\label{eq:holder_mid2}
    z_t^T \Delta^\dagger\Delta (\widetilde{P_t} - P_{0, t}) \\\nonumber&\leq ||(\Delta^\dagger)^T z_t||_\infty \left\|(\widetilde{P}_t - \widetilde{P}_{t-1}) + (P_{0, t} - P_{0, t-1})\right\|_1 \\\nonumber&\leq ||(\Delta^\dagger)^T z_t||_\infty \bigg(||\widetilde{P}_t - \widetilde{P}_{t-1}||_1 + ||P_{0, t} - P_{0, t-1}||_1\bigg)
\end{align}

Note $z_t$ are all independent, and thus their summation is treated as a constant. Thus, using equations ~\eqref{eq:holder_mid} and ~\eqref{eq:holder_mid2} and after applying summation, we have

$$\sumtot z_t^T \Delta^\dagger\Delta (\widetilde{P_t} - P_{0, t})\leqwithspaces \sumtot ||(\Delta^\dagger)^T z_t||_\infty \bigg(\sumtwotot ||\widetilde{P}_t - \widetilde{P}_{t-1}||_1 + \sumtwotot ||P_{0, t} - P_{0, t-1}||_1\bigg).$$

By picking the right value for $\lambda_1 \geq \frac 1 T \sumtot ||(\Delta^\dagger)^T z_t||_\infty$, then above equation can simplified as

$$\sumtot z_t^T \Delta^\dagger\Delta (\widetilde{P_t} - P_{0, t})\leqwithspaces \lambda_1 T \sumtwotot ||\widetilde{P}_t - \widetilde{P}_{t-1}||_1 + \lambda_1 T \sumtwotot ||P_{0, t} - P_{0, t-1}||_1.$$

Consequently, we use above inequality in ~\eqref{eq:before_holder}, and rewrite it as in the following

\begin{align*}
    \sumtot ||\widetilde{P_t} - P_{0, t}||_R^2 \\\leq\;\;& 2\lambda_1 T \sumtwotot ||\widetilde{P}_t - \widetilde{P}_{t-1}||_1 + 2\lambda_1 T \sumtwotot ||P_{0, t} - P_{0, t-1}||_1 \\&+ 2\lambda_1 T \sumtwotot ||P_{0, t} - P_{0, t-1}||_{1, 1} - 2\lambda_1 T \sumtwotot ||\widetilde{P_t} - \widetilde{P}_{t-1}||_{1, 1}.
\end{align*}

And thereupon

\begin{equation}\label{eq:final_inequality}
    \sumtot ||\widetilde{P_t} - P_{0, t}||_R^2 \leqwithspaces 4\lambda_1 T \sumtwotot ||P_{0, t} - P_{0, t-1}||_{1, 1}.
\end{equation}

Similar to previous studies~\cite{wang2016trend}, we know $||(\Delta^\dagger)^T z_t|| = \bigoprob(M\sqrt{\log r})$ by a standard result on the maximum of Gaussians (derived using the union bound, and Mills' bound on the Gaussian tail), where $M$ is the maximum $l2-$norm of the columns of $\Delta^\dagger$. Thus, for the hyperparameter $\lambda_1$ we know

$$\lambda_1 \geq \frac 1 T \sumtot ||(\Delta^\dagger)^T z_t||_\infty = \bigoprob(\frac M T\sqrt{\log r}).$$

And based on the total variation growing condition in ~\eqref{eq:growing_total_variation_condition} and the aforementioned choice of $\lambda_1$, inequality in ~\eqref{eq:final_inequality} gives the below inequality

\begin{equation}\label{eq:row_space}
    \sumtot ||\widetilde{P_t} - P_{0, t}||_R^2 \leqwithspaces 4 (\frac M T\sqrt{\log r}) T n^2 C_T = 4n^2 M C_T \sqrt{\log r}.
\end{equation}

The convergence rate for the entire problem, by using ~~\eqref{eq:row_space} and ~~\eqref{eq:null_space}, would be the big o probability of the following

\begin{align*}
    \sumtot ||\widetilde{P_t} - P_{0, t}||_F^2 &\leqwithspaces \sumtot ||\widetilde{P_t} - P_{0, t}||_{R_{\bot}}^2 + \sumtot ||\widetilde{P_t} - P_{0, t}||_R^2\\
    &\leqwithspaces n^2 \text{nullity}(\Delta) + 4n^2 M C_T \sqrt{\log r},
\end{align*}

And thus,

\begin{equation*}
    \sumtot ||\widetilde{P_t} - P_{0, t}||_F^2 = \bigoprob \bigg(n^2 (\text{nullity}(\Delta) + M \sqrt{\log r} C_T) \bigg) \qedhere
\end{equation*}

\end{proof}

\end{document}